%% file: ms.tex
\documentclass[12pt,preprint]{aastex}
\def\degr{\hbox{$^\circ$}}
\def\arcmin{\hbox{$^\prime$}}
\def\arcsec{\hbox{$^{\prime\prime}$}}

\def\fh{\hbox{$\!\!^{\rm h}$}}
\def\fm{\hbox{$\!\!^{\rm m}$}}
\def\fs{\hbox{$\!\!^{\rm s}$}}
\def\220{$^{13}$CO(2$\rightarrow$1)}
\def\460{$^{12}$CO(4$\rightarrow$3)}
\def\810{$^{12}$CO(7$\rightarrow$6)}
\def\kms{\ km s$^{-1}$}
\begin{document}

\title{AST/RO $^{13}$CO(J=2$\rightarrow$1) and $^{12}$CO(J=4$\rightarrow$3)
Mapping \\ of Southern Spitzer c2d Small Clouds and Cores}

\author{A. L\"ohr\altaffilmark{1,2}, T. L. Bourke\altaffilmark{1}, A. P.
Lane\altaffilmark{1}, P. C. Myers\altaffilmark{1}, \\ S. C.
Parshley\altaffilmark{3}, A. A. Stark\altaffilmark{1}, N. F. H.
Tothill\altaffilmark{1}}

\altaffiltext{1}{Harvard-Smithsonian Center for Astrophysics, 60 Garden
Street, Cambridge, MA 02138} 
\altaffiltext{2}{aloehr@cfa.harvard.edu}
\altaffiltext{3}{Cornell University, Ithaca, NY 14853}

\begin{abstract}
Forty molecular cloud cores in the southern hemisphere from the initial
Spitzer Space Telescope Cores-to-Disks (c2d) Legacy program source list
have been surveyed in \220, \460, and \810 with the Antarctic Submillimeter
Telescope and Remote Observatory (AST/RO).  The cores, ten of which contain
embedded sources, are located mostly in the Vela, Ophiuchus, Lupus,
Chamaeleon, Musca, and Scorpius complexes. \810 emission was undetected in
all 40 clouds. We present data of 40 sources in \220 and \460, significant
upper limits of \810, as well as a statistical analysis of the observed
properties of the clouds.  We find the typical \220 linewidth to be 2.0\kms
\ for cores with embedded stars, and 1.8\kms \ for all others.  The typical
\460 linewidth is 2.6 to 3.7\kms \ for cores with known embedded sources,
and 1.6 to 2.3\kms \ for all others.  The average $^{13}$CO column density
derived from the line intensities was found to be  1.9 $\times$
10$^{15}$~cm$^{-2}$ for cores with embedded stars, and 1.5 $\times$
10$^{15}$~cm$^{-2}$ for all others.  The average kinetic temperature in the
molecular cores, determined through a Large Velocity Gradient analysis of a
set of nine cores, has an average lower limit of 16 K and an average upper
limit of 26 K.  The average molecular hydrogen density has an average lower
limit of 10$^{2.9}$~cm$^{-3}$ and an average upper limit of
10$^{3.3}$~cm$^{-3}$ for all cores. For a different subset of nine cores,
we have derived masses. They range from 4 to 255 M$_{\odot}$.  Overall, our
c2d sample of southern molecular cores has a range of properties
(linewidth, column density, size, mass,
embedded stars) similar to those of past studies.  \end{abstract}


\section {INTRODUCTION}

Star formation occurs in dense molecular clouds and their dense cores are
the progenitors of protostars \citep{difrancesco, ward}. Such cores are
compact areas ($<$ 0.5 pc) of molecular gas at relatively high density ($>$
10$^{3}$ cm$^{-3}$).  Large efforts have been made in the observation and
characterization of such cores. One of the biggest and most recent of these
efforts is the Spitzer Space Telescope (SST) program `From Molecular Cores
to Planet-forming Disks' or `c2d' \citep{evans03}.  The c2d program
surveyed at infrared wavelengths over 100 relatively isolated cores that
span the evolutionary sequence from starless cores to protoplanetary disks
(Huard et al. 2006).  These sources include a wide range of cloud masses
and star-forming environments.  In order to understand how cores are formed
and subsequently evolve into stars, it is necessary to understand the gas
dynamics of the underlying molecular cloud.  The primary means of studying
molecular gas are observations of emission by the rotational transitions of
the ground vibrational state of carbon monoxide (CO). These spectral lines
occur at frequencies of $J \, \times$ 115 GHz, for transitions from the $J$
to the $J-1$ state.  A realistic picture of the thermodynamic state of a
molecular gas can be gained through observations of several transitions of
CO isotopes.  If the brightness of several of these lines is known, models
of the radiative transfer, such as a Large Velocity Gradient (LVG) model
\citep{goldr74}, can be used to determine density and temperature of the
gas. In order to avoid degeneracy in this solution, it is advisable to have
observations of a high-{\it{J}} transition line that is sufficiently high
in energy to be only weakly populated.

From the initial c2d source list of Evans et al. (2003), we have observed
with AST/RO all 40 molecular cores that are visible from the South Pole
($\delta \, < \, -16\degr$), in the \220, \460, and \810 transitions. (The
initial SST c2d source list was later shortened due to the time available
to the c2d program. Some of the sources in this paper have not been
observed with the SST.) Ten out of the 40 cores are known to have sources
embedded in them according to IRAS \citep{evans03,lee99}. The detection
threshold for star formation using IRAS is typically taken to be $\sim$0.1
L$_{\odot}$ (d/140\,pc) \citep{myers1987}. Cores in which IRAS has not
detected a source, which in this paper are referred to as `starless', may
still contain faint sources
\citep{young2004,tyla2005,tyla2006,huard2006,dunham2006}. The c2d team is
producing a list of candidate very low luminosity objects (VeLLOs) in all
cores.  The multiple transitions observed allow us to perform an LVG
analysis and gain information about the kinematic and thermodynamic
conditions in a large sample of molecular cores. These observations and
results can be used in conjunction with the c2d infrared results for
further studies of these cores.

In $\S$2, we describe the observatory, the observations, and the method of
data reduction. $\S$3 presents the dataset in the form of spectra, maps,
tables, and a statistical analysis. The Large Velocity Gradient analysis of
the radiative transfer is presented in $\S$3.1, which allows a
determination of kinetic temperature and molecular hydrogen density.
Masses for several cores are derived in $\S$3.2.  In $\S$4, we summarize
our results and present our conclusions.

\section{OBSERVATIONS}

The observations were performed during the austral winter of 2005 at the
Antarctic Submillimeter Telescope and Remote Observatory (AST/RO), which
was located at 2847 m elevation at the Amundsen-Scott South Pole Station.
These are among the final observations of AST/RO prior to its
decomissioning in December 2005.  Low water vapor and high atmospheric
stability make this site exceptionally good for submillimeter astronomy
\citep{cls1997,apl1998}.  AST/RO is a 1.7 m diameter, offset Gregorian
telescope that observes at wavelengths between 1.3 mm and 200 $\mu$m
\citep{aas2001,oberst06}. Emission from the 220.399 GHz $J=2\rightarrow 1$
transition of $^{13}$CO was mapped using an SIS receiver with 80 to 160 K
double-sideband receiver noise temperature \citep{kooi1992}.  A
dual-channel SIS waveguide receiver \citep{w1992,hon1997} was used for
simultaneous observation of 461 GHz and 807 GHz. Single-sideband receiver
noise temperatures were 340 to 440 K during observations of the 461.041 GHz
$J=4\rightarrow 3$ transition of $^{12}$CO and 700 to 1100 K for the
$J=7\rightarrow 6$ transition of $^{12}$CO at 806.652 GHz.

A multiple-position-switching mode was used with emission-free reference
positions at least 30\arcmin \ from the mapped area.  Integration times per
point were 347 seconds at 220 GHz and 220 seconds at 460/807 GHz.  The
beamsizes (FWHM) were 180\arcsec \ at 220 GHz, 103\arcsec \ at 460 GHz, and
58\arcsec \ at 807\,GHz. All maps were fully-sampled with neighbouring
points 1.5\arcmin \ apart at 220 GHz and 0.5\arcmin \ apart at 461/807 GHz.
Tests of the AST/RO pointing model indicate that pointing errors at the
time these data were taken could in some cases be as large as 2\arcmin \ at
220 GHz and 1\arcmin \ at 461/807 GHz.  

The telescope efficiency $\eta_{\ell}$ was estimated through the reduction
of a series of skydips to $\approx$ 81\% for the 220 GHz receiver,
$\approx$ 72\% at 461 GHz, and $\approx$ 80\% at 807 GHz. As described in
detail in Stark et al. (2001), the AST/RO main beam efficiency and
telescope efficiency are identical to the level that we can measure. The
telescope efficiency has been applied to the data as presented below.
Atmosphere-corrected system temperatures at the three frequencies ranged
from 280 to 480 K, 1000 to 2000 K, and 4000 to 10000 K, respectively.  Two
acousto-optical spectrometers (AOS) \citep{stw1989} with 670 kHz channel
width, 1.1 MHz resolution bandwidth, and 1.5 GHz total bandwidth were used
as a backend, resulting in velocity resolutions of 0.91\kms \ at 220 GHz,
0.65\kms \ at 461 GHz, and 0.37\kms \ at 807 GHz. For observationos at \220
we used a higher resolution AOS (HRAOS) \citep{stw1989} in addition to the
two mentioned above. The HRAOS has a channel width of 31.6\,kHz and a
resolution bandwidth of 64 MHz, resulting in a velocity resolution of
0.043\kms at 220 GHz.  The data from this spectrometer are used to derive
linewidths when the lower resolution was not sufficient and to confirm
linewidths close to the resolution limit at 220 GHz. These data are also
marked in the tables.  

A chopper wheel calibration technique was used to observe the sky and a
blackbody load of known temperature \citep{aas2001} every few minutes
during observations. Skydips were performed frequently to determine the
atmospheric transmission. To monitor the pointing accuracy, bright sources
(NGC\,3576 and/or the moon) were mapped every few hours. After regular
intervals or after changes to the receiver systems (tuning), the receiver
was manually calibrated against a liquid nitrogen load and a blackbody load
of known temperature. This process also corrects for the dark current of
the AOS's optical CCDs. The internal consistency of the AST/RO flux scale
was estimated from a large-scale map of Lupus, also taken in 2005 with
identical instrumental settings. This large map was constructed from
smaller maps, with some overlap between them, yielding over a thousand
points towards which more than one spectrum was measured at different
times. From these data, the flux scale is reproducable at the 30\% level
after correction for atmospheric absorption (Tothill et al., in
preparation).

The data taken for this survey were frequency calibrated in several ways.
Before and during observations, we observed a comb spectrum with every
automatic receiver calibration to monitor the frequency scale over the
entire observing period.  During measurements of the sky performed with
every automated calibration, the telescope observes a mesospheric telluric
line. The velocity scale of all observations was referenced to this scale,
resulting in an accuracy of plus/minus one channel.

The data were reduced using the COMB data reduction package as described in
\citet{aas2001}. Linear baselines were removed in all spectra at all
frequencies, excluding regions where emission was expected based on
previous $^{13}$CO($J=1\rightarrow 0$) observations (R. Otrupcek, private
communication; see also \citet{otru2000}).  Removal of first-order
baselines proved to be generally sufficient for the \220 and \460 data. In
very rare cases (0.01\% of all data), two or three Fourier components had
to be removed in addition to the linear baseline.  At \810 we did not
detect significant ($>$ 3\,$\sigma$) emission.  However, we were able to
deduce upper limits for the \810 intensity from our data for all of the
observed sources. Since we detected strong \810 emission in our pointing
sources, we can be certain that the non-detection of this line in the cores
is not a result of receiver failure. We conclude that the \810 emission
from the observed cores is so faint that it lies below the detection
sensitivity of our detectors. We thus deduce significant upper limits for
this emission.

A list of observed sources in order of increasing R.A., coordinates of the
map centers, as well as mapsizes is given in Table \ref{tab1}. Cores with
known embedded IRAS sources are marked ($\star$).  The molecular cores we
observed contain all cores from the source list of the Spitzer Space
Telescope c2d project \citep{evans03} that are visible from the South Pole
($\delta \, < \, -16\degr$).  The observed mapsizes were chosen to cover at
least the areas observed by c2d, but are larger in almost all cases.  \460
mapping was sometimes confined to regions where previously mapped \220
emission was observed.

\section{RESULTS AND ANALYSIS}

For each observed core, we present the highest intensity spectrum of 
%both detected transitions in Figures \ref{fig1a} - \ref{fig1e}.
both detected transitions in Figures 1a - 1e.  Observed and derived
parameters, and source information are listed in Table \ref{tab2} for the
\220 transition and in Table \ref{tab3} for the \460 transition. Molecular
cores with known embedded IRAS sources are marked with a ($\star$).  The
R.A. and Dec. coordinates in Tables \ref{tab2} and \ref{tab3} represent the
position of the spectrum with the maximum integrated intensity in each map.
The source velocity $v_{LSR}$ was obtained by fitting a Gaussian to the
line with maximum intensity. $\Delta v$ corresponds to the FWHM of this
fit.  In cases where the linewidth is lower or close to the spectrometer's
velocity resolution of 0.91\kms, we confirmed the linewidth from our higher
resolution data (0.043\kms). Cores for which we did this are marked in the
table.  $\int T_{\mathrm{MB}}(v)\,dv$ gives the intensity integrated over
the brightest line in the map, and rms is the root-mean-squared noise of
the channels in the spectrum.

In Table \ref{tab2}, the \220 intensity is converted to a total column
density of $^{13}$CO ($N_{^{13}CO}$) by assuming local thermodynamic
equilibrium (LTE):
\begin{equation} N_{^{13}CO}= 1.51 \times 10^{14} \ \frac{T_{ex} \
e^{\frac{5.3}{T}} \ \int \tau(v)\,dv}{1-e^{\frac{-10.6}{T_{ex}}}}
\end{equation}
Assuming optically thin \220 emission, that the source fills the main beam,
that the background temperature of 2.7 K can be neglected and that the
Rayleigh-Jeans correction can be neglected as well, we use the relation
\begin{equation} T_{ex} \ \tau(v) = T_{\mathrm{MB}}(v) \end{equation}
where $\tau(v)$ is the optical depth of the \220 line, T$_{\mathrm{MB}}$ is
the main beam brightness temperature and $T_{ex}$ is the excitation
temperature, which is usually taken to be the kinetic temperature in the
cloud: $T_{ex}$=T$_{kin}$=10 K (Up to $T_{ex}$ = 40 K the effect of this
parameter on the $^{13}$CO column density is less than a factor of 2).

In Figure \ref{fig2}, we show histograms of the linewidths of the two
detected transitions. The distribution of the \220 linewidth has a mean of
1.8\kms\ and a standard deviation of 0.4\kms . The cores with embedded
sources are represented by the black bars. It is noticeable that their
distribution is generally skewed to higher values. The mean of their
linewidths distribution is 2.0\kms and the standard deviation is 0.4\kms .
Cores with embedded stars tend to be warmer and more turbulent, and their
emission lines are therefore subject to more broadening, than are cores
without embedded stars.  If the gas kinetic temperature is typically 20 K,
the corresponding optically thin thermal linewidth of CO is only 0.2\kms,
so the observed widths are due primarily to turbulent motions and optical
depth effects.  DC253.8-10.9 shows an exceptionally narrow line with only
0.8\kms. It also is the weakest core in \220.  L328 has the broadest line
of the cores with no embedded IRAS source at 2.7\kms.  This core has
recently been found to have a VeLLO embedded (Lee, C.-W., in preparation).
The widest line at 2.9\kms \ is emitted by DC275.9+1.9, which does have an
embedded star \citep{tyla95}.

The distribution of the \460 linewidth also shows that the lines from cores
with embedded sources tend to be broader than lines from starless cores.
The mean linewidth of cores with embedded stars is 3.2\kms, with a standard
deviation of 1.3\kms. Cores with no embedded stars have a mean linewidth of
2.0\kms, with a standard deviation of 0.7\kms.  For this histogram, we did
not consider cores whose peak intensity is lower that 5$\sigma$. These
omitted cores are DC302.6-15.9, L63, CB68, and L328.  The widest line of
all cores is found in DC303.8-14.2 at 5.8\kms , which has an embedded star
\citep{tyla95}.  The widest line of cores with no embedded sources is
emitted by DC300.7-1.0 with 3.5\kms .

The survey of southern cores by \citet{vilas94} in $^{13}$CO(1$\rightarrow$
0) reports an average linewidth of 0.8 \kms. The only source observed in
both the Vilas-Boas et al. survey and this survey is CG30. \citet{vilas94}
gives a linewidth of 1.19\kms obtained with a resolution of 0.12\kms.
Using our high resolution spectrometer, we observed a linewidth of 1.2
$\pm$ 0.1\kms, which is in good agreement with \citet{vilas94}.

A histogram of the $^{13}$CO column density is shown in Figure \ref{fig3}.
The mean $^{13}$CO column density for cores with no embedded stars is 0.5
to 2.0 $\times$ 10$^{15}$~cm$^{-2}$ and slightly higher, around 1.0 to 2.0
$\times$ 10$^{15}$~cm$^{-2}$, for cores with embedded stars, which do not
show $^{13}$CO column densities lower than 1.0 $\times$
10$^{15}$~cm$^{-2}$.  Fifty-five percent of the cores have a $^{13}$CO
column density between 0.5 to 1.5 $\times$ 10$^{15}$~cm$^{-2}$, and 75\%
have column densities between 0.5 to 2.0 $\times$ 10$^{15}$~cm$^{-2}$.
Assuming a $^{13}$CO/H$_{2}$ ratio of 1.7 $\times$ 10$^{-6}$
\citep{tomwilson94,frerk82}, we can convert the $^{13}$CO column densities
to the H$_{2}$ column density.  The H$_{2}$ column density distribution has
a mean of 1.9 $\times$ 10$^{21}$~cm$^{-2}$ and a standard deviation of 0.7
$\times$ 10$^{21}$~cm$^{-2}$ for cores with embedded sources. For all other
sources the mean is 1.5 $\times$ 10$^{21}$~cm$^{-2}$ and the standard
deviation is 0.7 $\times$ 10$^{21}$~cm$^{-2}$.  Cores with embedded sources
tend to have higher column densities.  Seven of the observed cores lie in
Ophiuchus which has a high level of star formation activity
\citep{wilki92}.  In these cores the H$_{2}$ column density assumes high
values, with a mean of 3.7 $\times$ 10$^{21}$~cm$^{-2}$ and a dispersion of
1.5 $\times$ 10$^{21}$~cm$^{-2}$.  Twenty-three cores lie within clouds
where the star-forming activity is low (Vela, Musca, Coalsack, Chamaeleon
II and III, Lupus, Scorpius). These cores have a mean H$_{2}$ column
density of 2.7 $\times$ 10$^{21}$~cm$^{-2}$ with a dispersion of 1.0
$\times$ 10$^{21}$~cm$^{-2}$. This difference as an indicator of
star-forming activity was also observed by \citet{vilas2000} in a
$^{13}$CO(1-0) survey of different (none of the sources in their and our
survey overlap) dense condensations in dark clouds.  \citet{vilas2000} set
the limit for star-formation-enabling H$_{2}$ column density to as high as
1.1 $\times$ 10$^{22}$~cm$^{-2}$. This limit appears to be too high,
considering our results for B59.  B59 in Ophiuchus is the strongest core
observed and has the exceptionally high $^{13}$CO column density of 4.3
$\times$ 10$^{15}$~cm$^{-2}$ and H$_{2}$ column density of 7.3 $\times$
10$^{21}$~cm$^{-2}$, thus indicating strong star formation activity. A
group of embedded sources has already been found in this core
\citep{brooke}.

\subsection{Large Velocity Gradient Analysis}

For a more detailed analysis of our data, we chose only the cores most
suitable for the purpose: cores with strong ($>$ 5$\sigma$) emission in
both \220 and \460. We also made sure there was no large pointing offset
between intensity peaks at the two wavelengths.
%In Figures \ref{fig4a} - \ref{fig4e}, we present maps of the ten 
In Figures 4a to 4e, we present maps of the nine cores chosen for the LVG
analysis. The AST/RO contour maps are superimposed on optical images from
the Digital Sky Survey (DSS) retrieved from SkyView \citep{skyview}. The
cores are shown in order of ascending R.A. We present the majority of cores
superimposed on 15$\arcmin \times$ 15$\arcmin$ images, but AST/RO map sizes
vary.  Maps in the same row show the same source, the left column shows the
\220 emission, the right column shows the \460. All maps are labeled with
the source name, the observed transition, and the maximum integrated
intensity in K \kms. Contours in all maps are in percentage of the peak
intensities. We show contour levels in steps of 10\% of the highest
contour, whose value is given in each figure. The first maps on each page
show the AST/RO effective beamsizes of 216\arcsec \ at 220 GHz and
117\arcsec \ at 461 GHz.  Since the maps were made using Gaussian
interpolation, the effective beamsizes in the maps are slightly larger than
the actual telescope beamsizes. 

Maps that are larger than 15$\arcmin \times$ 15$\arcmin$ are shown in order
of ascending R.A. after the set of 15$\arcmin \times$ 15$\arcmin$ maps. The
top figure shows the \220 map, the bottom figure shows the \460 map.  In
cases with more than one source per map, the sources are labeled
individually. 

We can estimate the kinetic temperature $T_{kin}$, and the number density
of molecular hydrogen $n(H{_2})$, using a Large Velocity Gradient (LVG)
analysis of the radiative transfer \citep{goldr74}.  Our LVG radiative
transfer code has been developed and applied by M. Yan and S. Kim
\citep{skim} and has been implemented as a java applet
\footnote{{\tt{http://arcsec.sejong.ac.kr/$\sim$skim/lvg2.html}}}.  This
model simulates a plane-parallel cloud geometry and uses the CO collisional
rates from \citet{turner95}, newly-derived values for the H$_{2}$
ortho-to-para ratio ($\approx$ 2), and the collisional quenching rate of CO
by H$_{2}$ impact \citep{balakrishna02}.  The model has two input
parameters: the ratio of $^{12}$CO to $^{13}$CO abundance, and the ratio
$X$(CO)/$\nabla V$, where $X$(CO) is the fractional CO abundance and
$\nabla V$ denotes the velocity gradient. The abundance ratio
$^{12}$CO/$^{13}$CO is taken to be 60 in dense molecular clouds
\citep{tomwilson94}.  The $^{12}$CO/H$_{2}$ ratio is taken to be 10$^{-4}$
\citep{frerk82} (leading to a $^{13}$CO to H$_{2}$ ratio of 1.7 $\times$
10$^{-6}$) and the velocity gradient across the cores 1 \kms \ pc$^{-1}$. 

The \460 data were convolved with a Gaussian kernel to the lower \220
resolution.  The resulting peak intensities are listed in Table \ref{tab4}. 

Although we were unable to detect any significant \810 emission, we can use
these data to derive a significant upper limit for this emission for each
source. We convolved the higher resolution \810 data with a Gaussian kernel
to the \220 resolution and derived a 3$\sigma$ upper limit for the emission
through this relation:
\begin{equation} \rm{3  \sigma =3 \times \frac{rms \times \sqrt{N} \times
channelwidth \times 0.94}{FWHM(4-3)}} \end{equation}
where `rms' is the rms noise of the co-added spectrum. N is the number of
channels over which the rms was measured (twice the expected linewidth of
the \810 line).  `Channelwidth' is the width of the acousto-optic
spectrometer channels at this wavelength.  FWHM(4-3) is the half-power
width of the corresponding \460 line.  The factor of 0.94 comes from the
relation for Gaussian profiles: FWHM $\times$ peak = 0.94 $\times$ area.

The results are shown in Table \ref{tab4}, where upper and lower limits are
denoted u.l. and l.l.  Since the line brightness temperature
T$^{12}_{7\rightarrow 6}$ of the \810 line is an upper limit, the ratio
$T^{12}_{7\rightarrow6}$/$T^{12}_{4\rightarrow 3}$, the kinetic temperature
and the H$_{2}$ density are also upper limits.  To obtain the lower
temperature and density limits, we assume no \810 emission at all:
T$^{12}_{7\rightarrow 6}$ = 0.

As can be seen from Table \ref{tab4}, the lower temperature limits average
to 16 K and are not greater than 20 K for any source.  The distribution of
the upper temperature limits have an average of 26 K.  The H$_{2}$ density
distribution is fairly uniform, ranging from 10$^{2.7}$~cm$^{-3}$ to
10$^{3.7}$~cm$^{-3}$ with an average lower limit of 10$^{2.9}$~cm$^{-3}$
and an average upper limit of 10$^{3.3}$~cm$^{-3}$ for all sources.  Due to
the small sample size, we cannot detect a significant difference in the
mean densities of cores with and without embedded stars.

\subsection{Masses}

Nine cores were chosen for the derivation of their masses. We selected
cores whose half-peak contours in \220 are closed to derive the cores'
masses.  These cores are also resolved. Maps of the selected sources are
presented in Figures 4 and \ref{fig5}. We show contour levels in steps of
10\% of the highest contour, whose value is given in each figure.
DC302.6-15.9 and DC303.8-14.2 are at very high elevation (79\degr \ and
77\degr) and possibly suffer from a systematic pointing offset as suggested
from the background image. For the calculation of their masses, a potential
pointing offset is irrelevant. There is no evidence that other cores show a
similar pointing offset.  We can obtain the total mass of the core using 
\begin{equation} M=\langle m \rangle \int_{A} N(H_{2})\ dA \end{equation}
where $A$ is the projected area and $\langle m \rangle$ is the mean mass
per hydrogen molecule. This mean mass per hydrogen molecule $\langle m
\rangle$ accounts for a helium abundance of one helium atom per five
hydrogen molecules.  We take into account that the sources are not
spherical, but elliptical, and list their two radii (major and minor axes)
in Table \ref{tab5}. We derive an average $^{13}$CO column density for the
core area that is defined by the 50\% peak intensity contour by co-adding
all data that were taken in this area and making use of equation (1). This
conversion assumes an excitation temperature of $T_{ex}$ = 10 K. Up to
$T_{ex}$ = 40 K, the effect of this parameter on the $^{13}$CO column
density is less than a factor of 2.  By assuming a ratio of 1.7 $\times$
10$^{-6}$ between $^{13}$CO and H$_{2}$ \citep{tomwilson94,frerk82}, we
then convert the $^{13}$CO column densities to the H$_{2}$ column densities
listed in Table \ref{tab5} and use the relation above to determine their
masses.  Their distances are quoted from the literature: DC259.5-16.4,
CG30, DC267.4-7.5, DC253.6+2.0, and DC274.2-0.4 lie within the Vela complex
at 450 $\pm$ 50 pc \citep{woermann}; DC297.7-2.8 is associated with the
Coalsack at 150 $\pm$ 30 pc \citep{franco, corradi}; DC302.6-15.9 and
DC303.8-14.2 lie in the Chamaeleon cloud complexes at distances between 150
- 180 pc \citep{whittet,knude}; L100 lies within Ophiuchus at 125 $\pm$ 25
pc \citep{geus}.  The results are listed in Table \ref{tab5}. If the upper
limit kinetic temperature was obtained through the LVG analysis, we list it
again. 

The derived masses range from 4 to 255 M$_{\odot}$, depending on the size
of the core. The highest-mass core, DC267.4-7.5, is also the largest.  The
smallest and weakest core, DC302.6-15.9, does have the lowest mass.  A
correlation between temperature and mass cannot be established. 

We have found three cores whose masses have been derived from different
observations.  \citet{vilas94} derived the mass of CG30 to be 43
M$_{\odot}$ with an uncertainty of a factor of 2.5, but they also quote a
shorter distance of 300 pc. With a distance of 450 pc, this mass would be
97 M$_{\odot}$, which is within a factor of two of our value (58
M$_{\odot}$).  \citet{tyla97} have calculated the mass of DC297.7-2.8 to be
40 M$_{\odot}$, using a larger radius and larger distance for the source.
With the distance we use, 150 pc, this value becomes 23 M$_{\odot}$, which
is also within a factor of two of our value (12 M$_{\odot}$).
\citet{keto86} calculated the mass of DC259.5-16.4 to be 41 M$_{\odot}$,
but they used the much smaller distance of 100 pc. Combining their
observations with a distance of 450 pc leads to a mass for DC259.5-16.4 of
830 M$_{\odot}$.

\section{SUMMARY AND CONCLUSIONS}

In order to study the physical conditions in small isolated molecular
clouds, we have mapped 40 cores in the southern hemisphere in \220, \460,
and \810 with AST/RO. The sources cover all cores visible from the South
Pole ($\delta \, < \, -16\degr$) from the initial c2d source list
\citep{evans03}.  The AST/RO survey at millimeter and submillimeter
wavelengths provides a direct diagnostic of the dominant gas components in
the clouds. By observing the $^{12}$CO and $^{13}$CO lines, we have
obtained information about the physical conditions within the cloud.
$^{13}$CO is a tracer of moderately dense ($\sim$ 10$^{3}$ cm$^{-3}$) gas.
It is generally optically thin and hence is a good tracer of column
density. It probes the outer regions of dense cores (10 to 20 K), but not
the densest interiors where a small percentage of the total $^{13}$CO mass
is depleted within a small percentage of the total size ($\sim$ 10$^{3}$
AU; \citet{tafalla}).  The most important submillimeter measurement is the
\460 transition: this line is the lowest-lying of the mid-{\it{J}} CO lines
that constrain density rather strongly. These data have been used to
determine meaningful conclusions about the physical properties of the
molecular cores.  The detection of the \810 transition was unlikely, since
the molecular cores are rather cold. From our data, we were able to put
upper limits on the intensities in this transition, which were important
for a Large Velocity Gradient (LVG) analysis. 

We find typical \220 linewidths to be 1.8 to 2.3\kms \ for cores with
embedded stars and 1.6 to 1.9\kms for all others.  The slightly higher
values for sources with embedded stars result from warmer gas whose
emission is more subject to thermal and turbulent broadening of the lines.
Only 10 out of 40 cores have linewidths outside the 1.5 to 2.1\kms \
window, the smallest being 0.8\kms \ and the largest being 2.7\kms. The
largest linewidth is found in DC275.9+1.9, a core with an embedded star;
the smallest is found in the rather weak source DC253.8-10.9, with no signs
of star formation. 

The distribution of the \460 linewidth is much broader. Typical values lie
in the range of 1.6 to 2.3\kms \ for cores with no embedded sources, and
2.6 to 3.7\kms \ for cores with embedded stars. The distribution of their
linewidths is rather flat, with the largest linewidth of 5.8\kms \
occurring in DC303.8-14.2.  From the $^{13}$CO line intensities, we were
able to derive the $^{13}$CO column density and the H$_{2}$ column density
($\S$ 3).  The mean of the $^{13}$CO column density is found to be 1.9
$\times$ 10$^{15}$~cm$^{-2}$ for cores with embedded stars, and 1.5
$\times$ 10$^{15}$~cm$^{-2}$ for all others.  Due to its exceptionally
strong \220 line intensity, B59 also stands out here with a $^{13}$CO
column density of 4.3 $\times$ 10$^{15}$~cm$^{-2}$ and an H$_{2}$ column
density of 7.3 $\times$ 10$^{21}$~cm$^{-2}$.

For nine cores, we determined the kinetic temperature and molecular
hydrogen density through an LVG analysis.  For the \810 transition, we
determined upper intensity limits, since we have no significant detection
of this  line. As a result, all temperatures and densities are also upper
limits. In addition, we give a lower temperature limit, assuming the
absence of \810 emission.  For these nine cores, four of which have
embedded stars, we find an average lower temperature limit of 16 K and an
average upper limit of 26 K. The lower temperature limit is smaller than 21
K for all cores.  The average upper limit on the molecular hydrogen density
was found to be 10$^{3.3}$~cm$^{-3}$ for all cores. The average lower limit
on the molecular hydrogen density was found to be 10$^{2.9}$~cm$^{-3}$. Due
to the small sample size, we cannot detect a significant difference between
the mean densities of cores with and without embedded stars.

We have obtained masses for nine cores. They range from 4 to 255
M$_{\odot}$, depending on the size of the core.  A correlation between
temperature or density and mass cannot be established, which is mainly due
to the small sample size. 

In general, the sources in this survey behave rather uniformly with only
very few exceptions.  Cores with embedded sources tend to have larger
linewidths and higher $^{13}$CO column density than cores with no embedded
stars.

From the comparison to previously published CO surveys of molecular cores
\citep{keto86,vilas94,vilas2000}, we can say that the c2d sample of
molecular cores has a range of properties (linewidth, column density, size,
mass, embedded stars) similar to those of past studies. Only three sources
from the sample covered in this survey have previously been observed in
other CO surveys, albeit in lower rotational transitions.  To obtain even
more accurate results with an LVG analysis, a survey of these sources at
one other CO transition would be very valuable.  

\acknowledgements

The authors would like to thank Jacob Kooi of Caltech for his excellent
work on the telescope and the receivers.  We thank the receiver group at
the University of Arizona and Greg Wright of Antiope Associates for their
work on the instrumentation; R. Schieder, J. Stutzki, and colleagues at the
University of K\"oln for their AOS's.  We are grateful to S. Kim and M. Yan
for making their LVG code available to us.  We acknowledge the use of
NASA's SkyView facility (\url{http://skyview.gsfc.nasa.gov}) located at
NASA Goddard Space Flight Center. This research was supported by the United
States National Science Foundation Office of Polar Programs, grants
OPP-0126090 and OPP-0441756.

\clearpage

\input{tab1.tex}
\input{tab2.tex}
\input{tab3.tex}
\input{tab4.tex}
\input{tab5.tex}

\clearpage

\begin{figure}
\figurenum{1a}
\label{fig1a}
\caption{\220 (left) and \460 (right) spectra with the highest integrated
intensities.  Coordinates of the spectra are given in Tables \ref{tab2} and
\ref{tab3}.}
\vspace*{5mm}
\epsscale{0.85}
\plotone{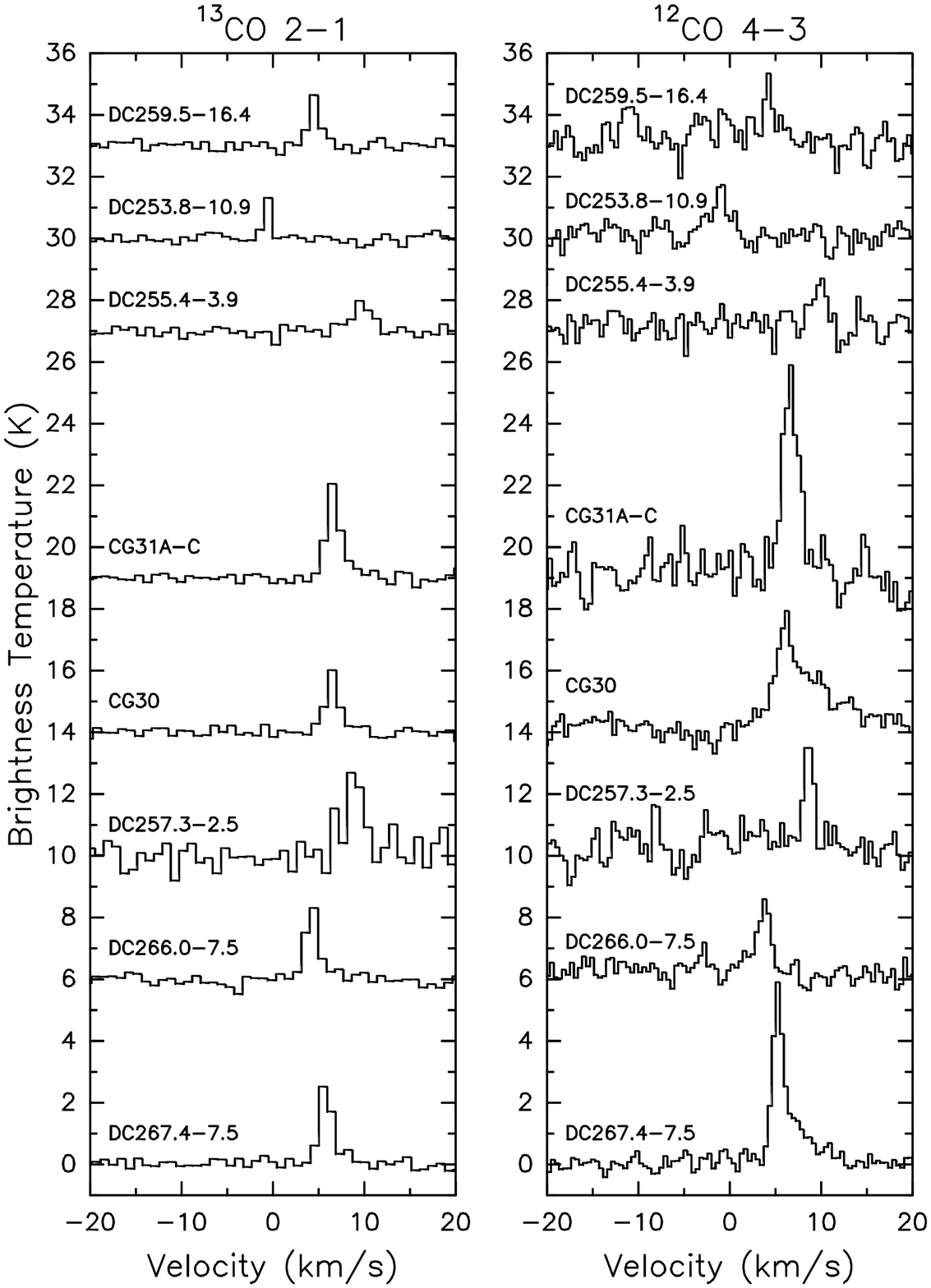}
\end{figure}
\clearpage
{\plotone{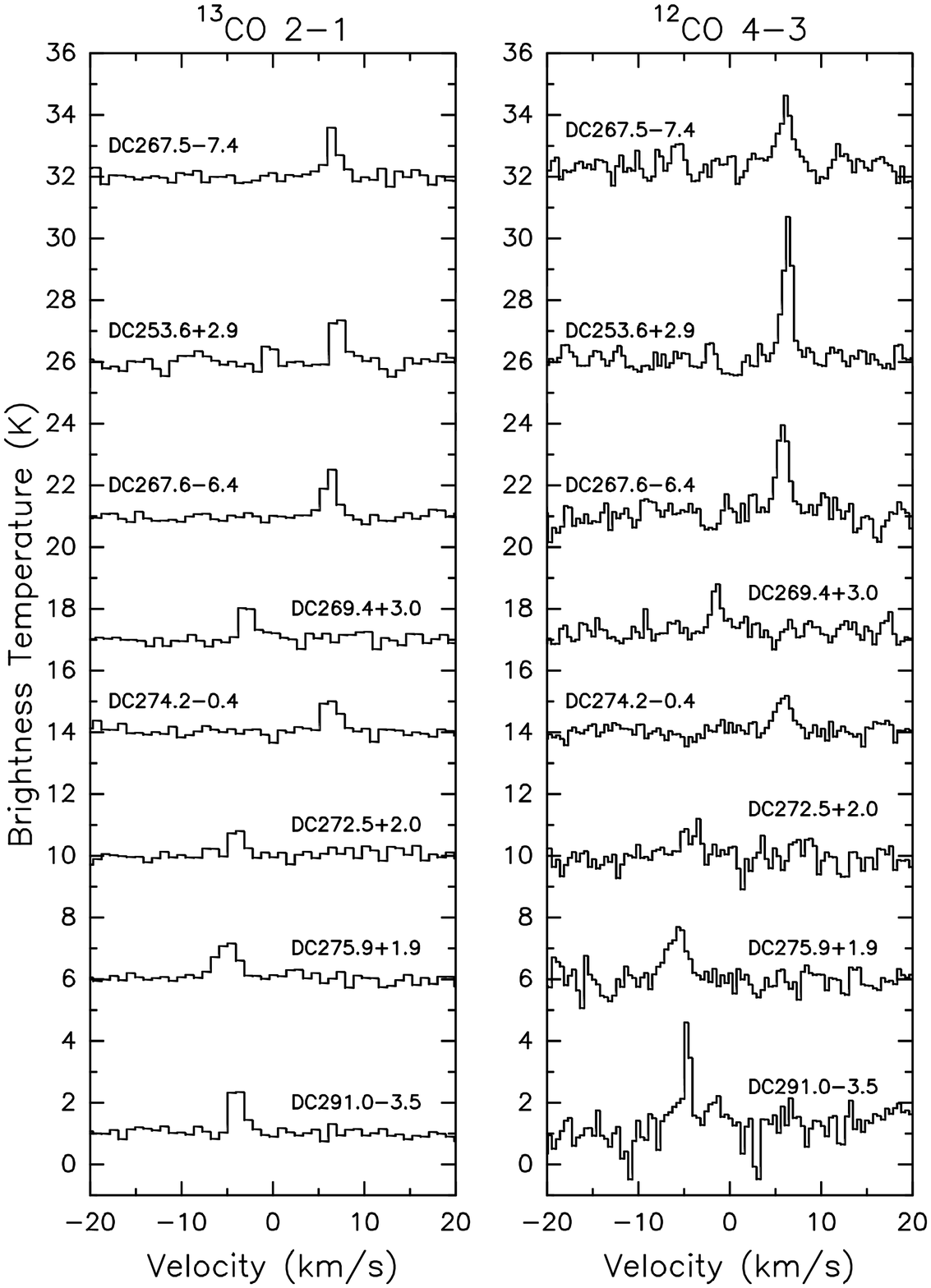}}\\[5mm]
\centerline{Fig. 1b. ---}
\clearpage
{\plotone{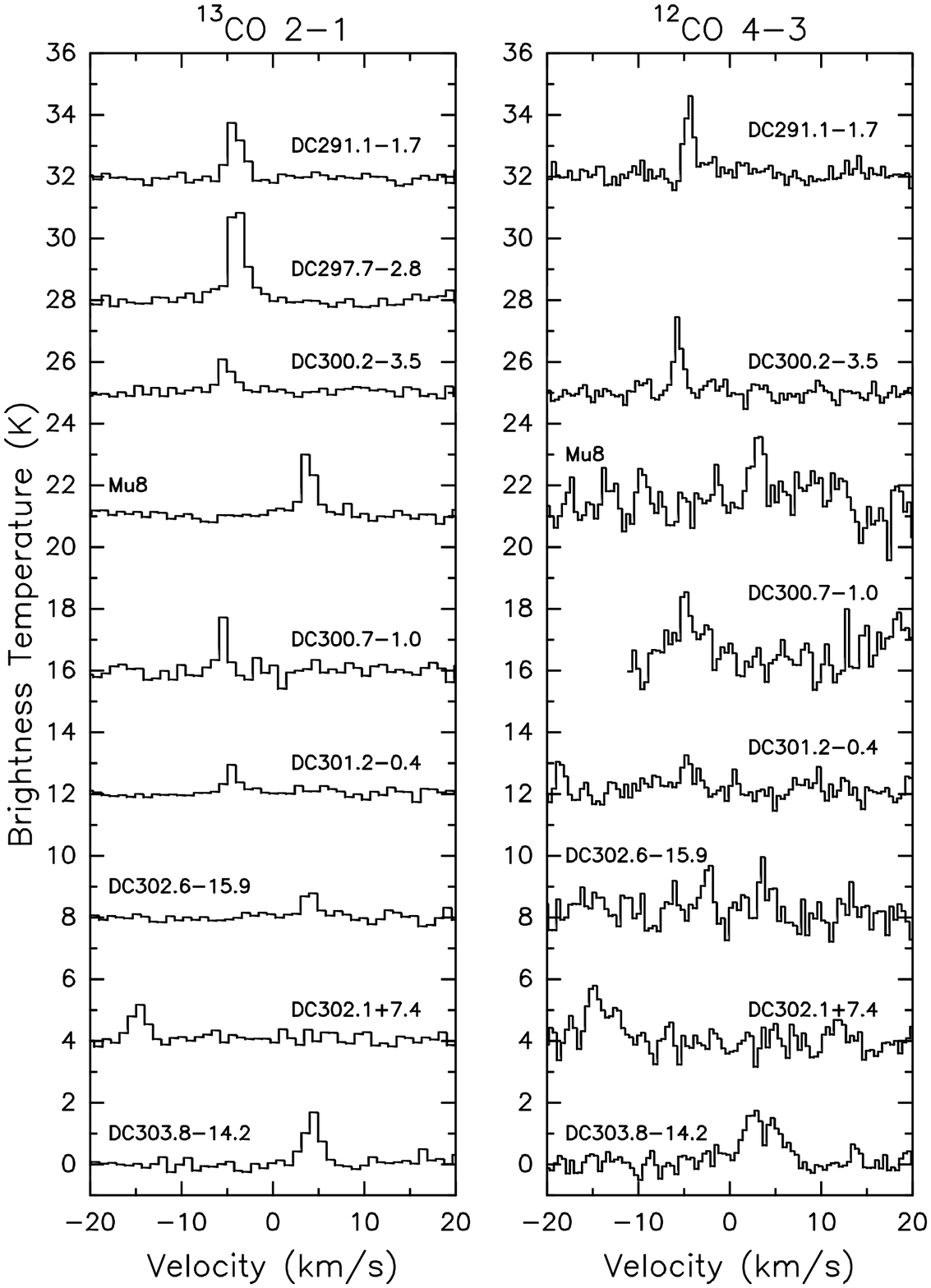}}\\[5mm]
\centerline{Fig. 1c. ---}
\clearpage
{\plotone{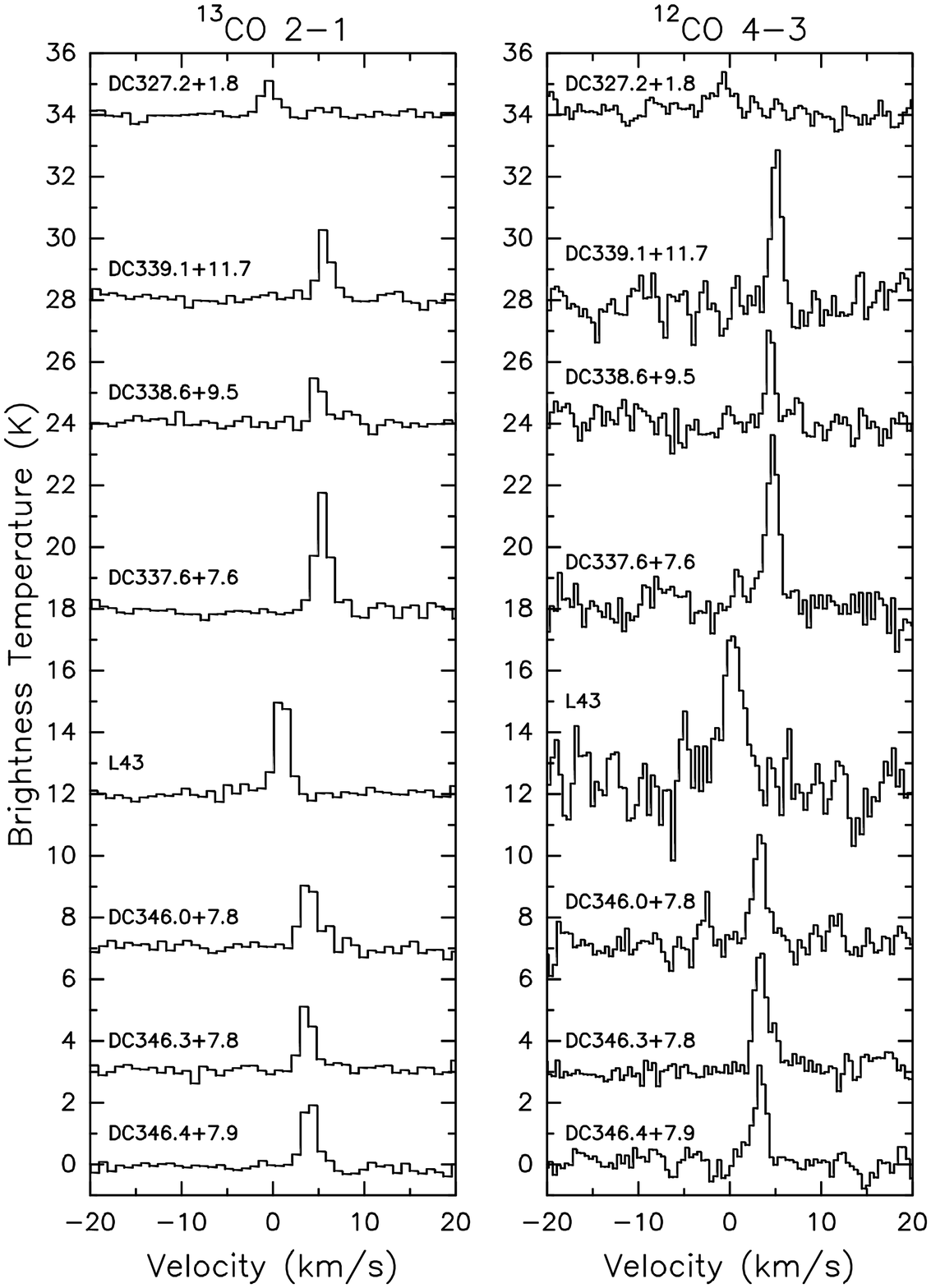}}\\[5mm]
\centerline{Fig. 1d. ---}
\clearpage
{\plotone{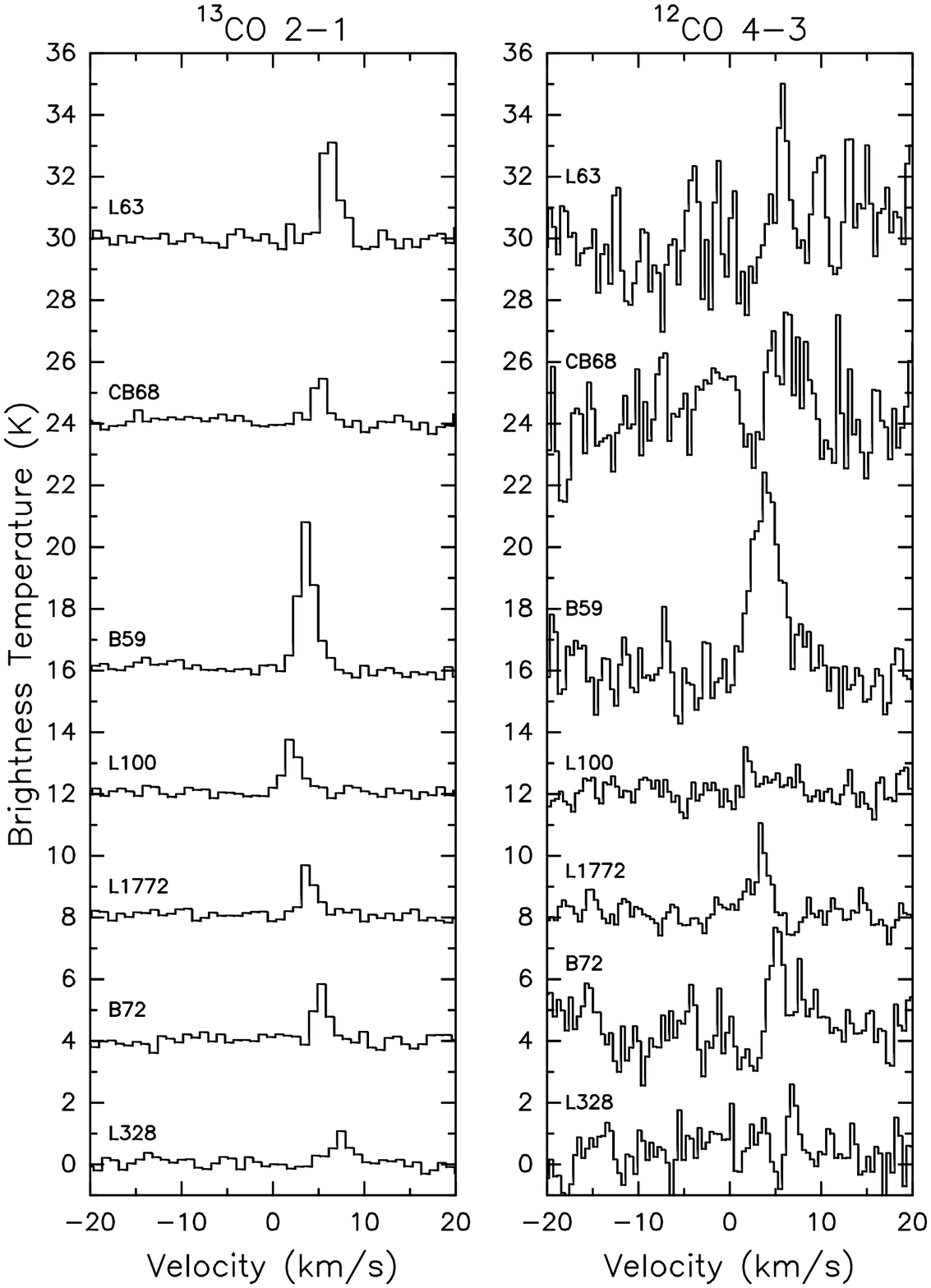}}\\[5mm]
\centerline{Fig. 1e. ---}

\begin{figure}[h!]
\figurenum{2}
\label{fig2}
\caption{Histograms of the \220 linewidth (left) and 
\460 (right). Grey bars represent sources
with no embedded stars, black bars represent sources with embedded IRAS sources.}\vspace*{5mm}
\epsscale{0.8}
\plottwo{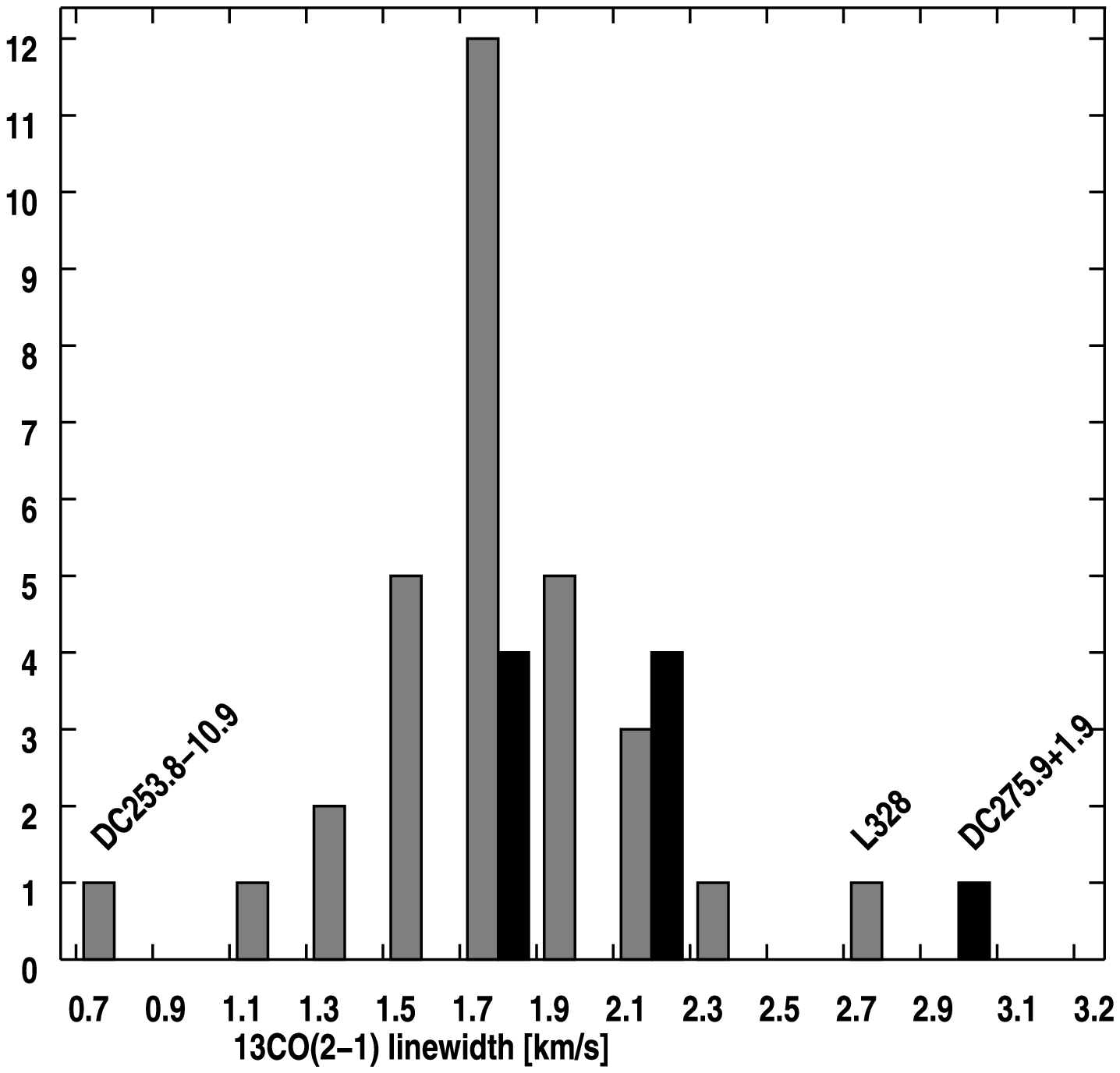}{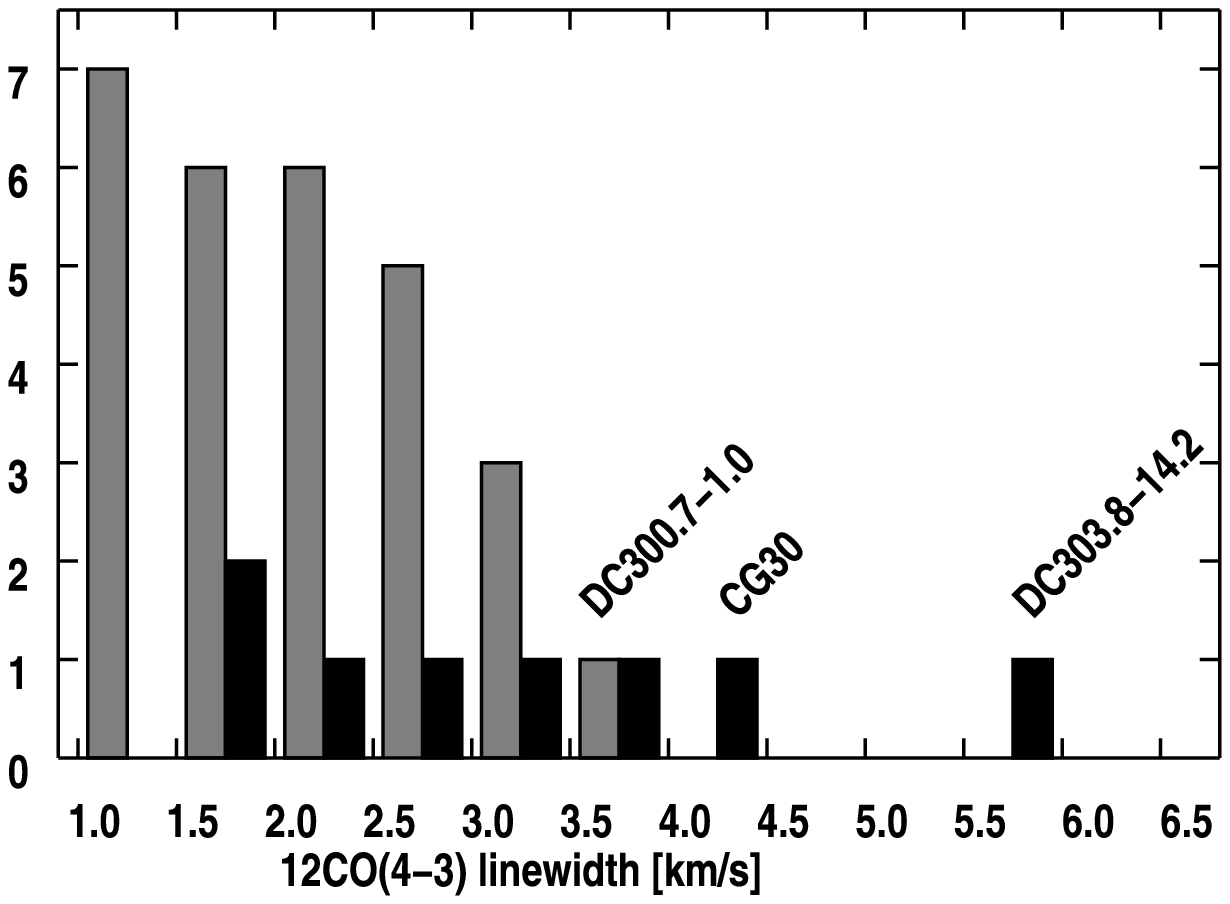}
\end{figure}

\begin{figure}[h!]
\figurenum{3}
\label{fig3}
\caption{Histogram of the $^{13}$CO 
column density for all sources. Grey bars represent sources
with no embedded stars, black bars represent sources with embedded IRAS sources.}\vspace*{5mm}
\epsscale{0.4}
\plotone{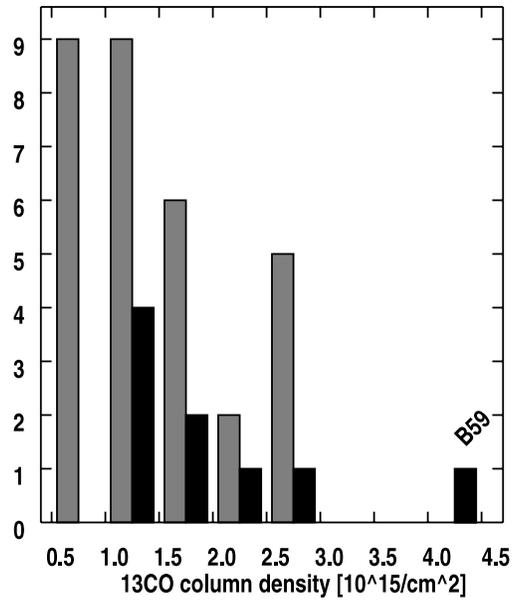}
\end{figure}

\begin{figure}[h!]
\figurenum{4a}
\label{fig4a}
\caption{Maps in 2 transitions of sources chosen for LVG analysis:
\220 (left/top) and \460 (right/bottom). Contours are in steps of 10\% of 
the peak intensity, listed in each map. The effective beamsizes are indicated
by the circles and are 216\arcsec \ in \220 and 117\arcsec \ in \460.
 (a) DC259.5-16.4, DC274.2-0.4, DC339.1+11.7, (b) L100, 
(c) CG31A-C, CG30, (d) DC267.5-7.4, DC267.4-7.5, (e) B59}\vspace*{5mm}
\epsscale{1.}
\plottwo{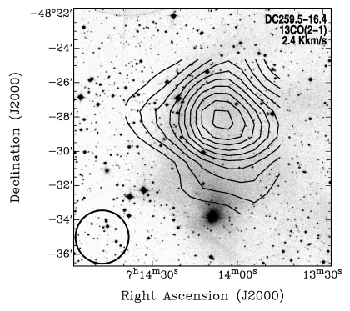}{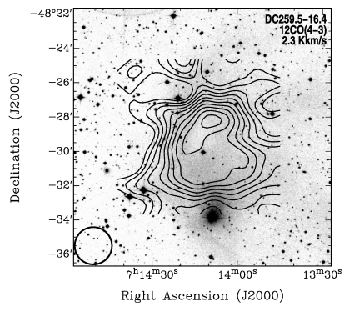}\\
\epsscale{1.}
\plottwo{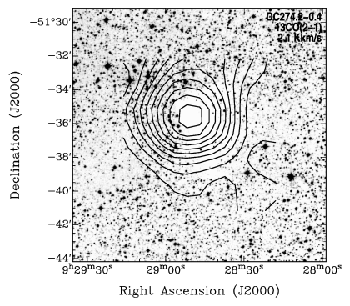}{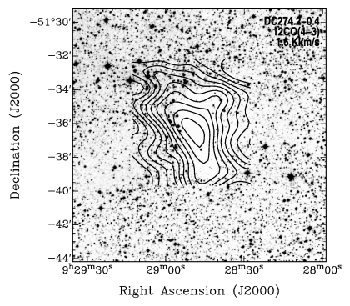}\\
\epsscale{1.}
\plottwo{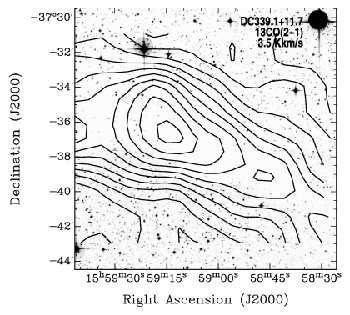}{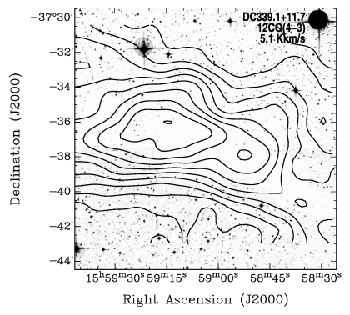}
\end{figure}
\clearpage
\epsscale{1.2}
{\plottwo{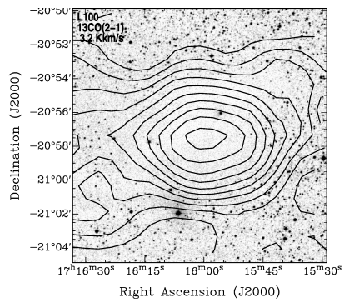}{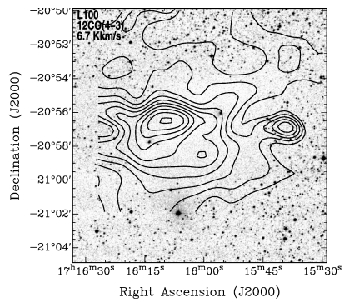}}\\
\clearpage
\epsscale{0.65}
\plotone{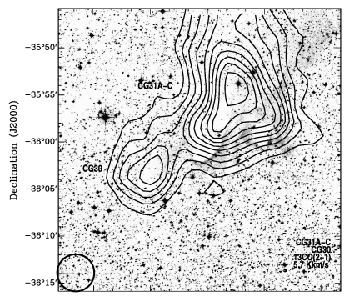}
\epsscale{0.65}
\plotone{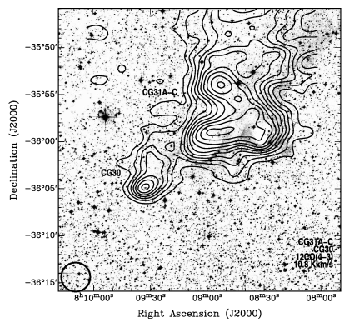}\\[5mm]
\centerline{Fig. 4c. ---}
\clearpage
\epsscale{0.85}
\plotone{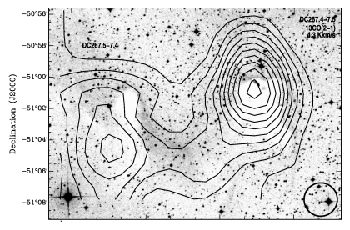}
\epsscale{0.85}
\plotone{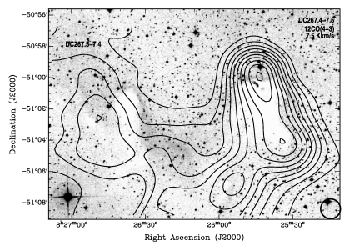}\\[5mm]
\centerline{Fig. 4d. ---}
\clearpage
\epsscale{0.65}
{\plotone{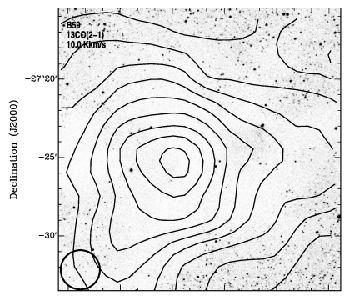}}
\epsscale{0.65}
{\plotone{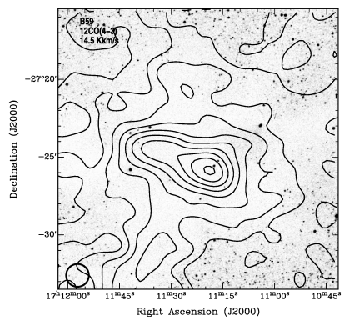}}\\[5mm]
\centerline{Fig. 4e. ---}

\clearpage
\begin{figure}[h!]
\figurenum{5}
\label{fig5}
\caption{Maps of DC253.6+2.0, DC297.7-2.8, DC302.6-15.9, and DC303.8-14.2 in
\220. The effective beamsize of 216\arcsec \ is indicated by the circle.}\vspace*{5mm}
\epsscale{1.2}
\plottwo{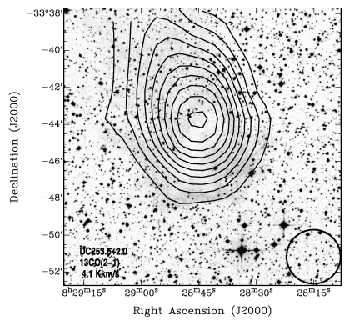}{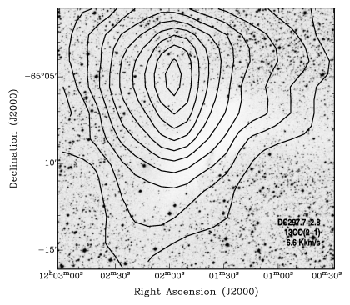}\\
\epsscale{1.2}
\plottwo{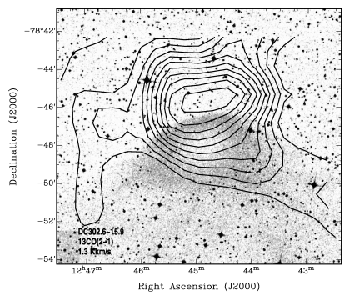}{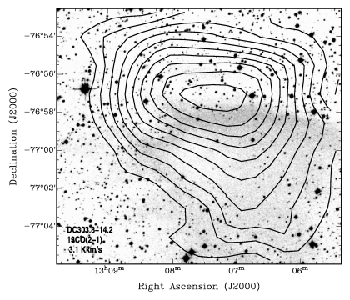}
\end{figure}

\end{document}

%% file: tab1.tex
\begin{deluxetable}{lrrrcc}
\tablewidth{0pc}
\tablenum{1}
\tablecaption{Source list.\label{tab1}}
\tablehead{
\colhead{Source name} &\colhead{Other names}& \colhead{$\alpha$ [J2000]}\tablenotemark{1}& \colhead{$\delta$ [J2000]}\tablenotemark{1} 
& \colhead{$^{13}$CO(J=2$\rightarrow$1)}& \colhead{$^{12}$CO(J=4$\rightarrow$3)}  \\ 
 \phn & \phn & \colhead{[\ \fh \ \ \fm \ \ \fs]}  &  \colhead{[\degr \ \arcmin \
   \arcsec]} &\colhead{$\alpha$[\arcmin]$\times \delta$[\arcmin]} & \colhead{$\alpha$[\arcmin]$\times \delta$[\arcmin]}}
\startdata
DC259.5-16.4 & BHR22 & 07 14 12.2 & -48 29 10 & 9 $\times$ 9 &  9 $\times$ 9 \\
DC253.8-10.9 & BHR14 & 07 29 32.7 & -41 10 30 & 9 $\times$ 9 &  9 $\times$ 9 \\
DC255.4-3.9 & BHR16 & 08 05 26.3 & -39 08 56 & 9 $\times$ 9 &  9 $\times$ 9 \\
$\star$\tablenotemark{2} CG30, CG31A-C & BHR12, BHR9-11 & 08 09 04.0 & -36 00 51 &  30 $\times$ 30 & 24.5  $\times$ 23 \\
DC257.3-2.5 & - & 08 17 09.3 & -39 52 25 & 9 $\times$ 15 & 9.5  $\times$ 15 \\ 
DC266.0-7.5 & - & 08 21 22.9 & -49 50 47 & 9 $\times$ 9 & 9 $\times$ 9 \\ 
$\star$ DC267.4-7.5 & BHR36 & 08 25 48.3 & -51 01 18 & 12 $\times$ 15 & 11.5 $\times$ 12 \\
DC267.5-7.4 & BHR37 & 08 26 37.0 & -51 02 31 & 9 $\times$ 9 & 9  $\times$ 9 \\ 
DC253.6+2.9 & BHR13 & 08 28 44.0 &  -33 45 12 & 15 $\times$ 15 & 14.5  $\times$ 15 \\ 
DC267.6-6.4 & BHR40 & 08 32 00.8 & -50 32 51  & 15 $\times$ 15 & 9  $\times$ 9 \\ 
DC269.4+3.0 & BHR43 & 09 22 22.0 & -45 50 25 & 12 $\times$ 9 & 12.5  $\times$ 9 \\
DC274.2-0.4 & BHR53 & 09 28 47.0 & -51 36 40 & 9 $\times$ 9 & 7  $\times$ 7.5 \\
DC272.5+2.0 & BHR47 & 09 31 02.8 & -48 38 22 & 9 $\times$ 9 & 11  $\times$ 6.5 \\
$\star$ DC275.9+1.9 & BHR55 & 09 46 46.0 & -51 05 12 & 9 $\times$ 9 & 9.5  $\times$ 6.5 \\ 
DC291.0-3.5 & - & 10 59 43.0 & -63 43 43 & 15 $\times$ 15 & 15 $\times$ 15 \\ 
DC291.1-1.7 & BHR59 & 11 06 52.0 & -62 06 43 & 9 $\times$ 9 & 9 $\times$ 9 \\
$\star$ DC297.7-2.8 & BHR71 & 12 01 35.9 & -65 08 37 & 15 $\times$ 15 & - \tablenotemark{3} \\
DC300.2-3.5 & BHR75 & 12 24 19.0 & -66 11 52 & 9 $\times$ 9 & 11.5  $\times$ 9 \\
Mu8 & Musca Dark Cloud & 12 29 35.0 & -71 09 35 & 15 $\times$ 21 & 15 $\times$ 21 \\ 
DC300.7-1.0 & BHR77 & 12 31 34.0 &  -63 44 32  & 15 $\times$ 15 & 15 $\times$ 15 \\ 
DC301.2-0.4 & BHR78 & 12 36 15.0 & -63 11 39 & 15 $\times$ 15 & 9 $\times$ 9 \\
DC302.6-15.9 & BHR84 & 12 44 57.6 & -78 48 21  & 15 $\times$ 12 & 15 $\times$ 9.5 \\ 
$\star$ DC302.1+7.4 & BHR83 & 12 45 38.5 & -55 25 18 & 9 $\times$ 9 & 7.5 $\times$ 9 \\ 
$\star$ DC303.8-14.2 & BHR86 & 13 07 35.7 & -77 00 05 & 15 $\times$ 15 & 12 $\times$ 15 \\ 
DC327.2+1.8 & BHR111 & 15 42 19.2 & -52 48 09  & 9 $\times$ 9 & 9 $\times$ 9 \\ 
DC339.1+11.7 & BHR129 &15 59 03.6 & -37 36 57 & 15 $\times$ 12 & 15 $\times$ 12.5 \\ 
DC338.6+9.5 & BHR126 & 16 04 28.0 & -39 37 56 & 9 $\times$ 9 & 9.5 $\times$ 7.5 \\
DC337.6+7.6 & - & 16 08 21.0 &  -41 42 19 & 18 $\times$ 18 & 17.5 $\times$ 18 \\ 
$\star$ L43 & RNO91 & 16 34 36.2 & -15 46 57 & 15 $\times$ 15 & 15 $\times$ 15 \\
DC346.0+7.8 & B231 & 16 36 54.7 & -35 36 08 & 9 $\times$ 12 & 9.5 $\times$ 12 \\
DC346.3+7.8 & - & 16 37 44.9 & -35 27 53 & 9 $\times$ 9 & 9.5 $\times$ 7.5 \\ 
DC346.4+7.9 & BHR144 & 16 37 33.0 &  -35 14 11 & 9 $\times$ 9 & 9 $\times$ 9 \\ 
L63 & - & 16 50 11.0 & -18 04 34 & 15 $\times$ 15 & 15 $\times$ 12.5 \\ 
$\star$ CB68 & L146 & 16 57 20.5 & -16 09 02 & 15 $\times$ 12 & 15 $\times$ 12 \\ 
$\star$ B59 & L1746 & 17 11 22.1  & -27 24 28  & 21 $\times$ 18 & 20.5 $\times$ 18 \\
$\star$ L100 & B62 & 17 16 01.0  & -20 57 21 & 15 $\times$ 15 & 14.5 $\times$ 12 \\
L1772 & B65 & 17 19 32.7 & -26 44 40 & 15 $\times$ 15 & 15 $\times$ 15 \\
B72 & CB83 & 17 23 45.0 & -23 41 28 & 18 $\times$ 12 & 17.5 $\times$ 12.5 \\
L328 & CB131 & 18 17 00.0 & -18 01 54 & 9 $\times$ 9 & 8 $\times$ 7.5 
\enddata
\tablenotetext{1}{\ Coordinates of the map center}
\tablenotetext{2}{\ Cores with known embedded sources as detected using IRAS are
  marked ($\star$).}
\tablenotetext{3}{\ \460 data on DC297.7-2.8 will be presented by Bourke et al., 
in preparation}
\end{deluxetable}

%% file: tab2.tex
\begin{deluxetable}{lrrrrrrr}
\tablewidth{0pc}
\tablenum{2}
\rotate
\tablecaption{Observed and derived parameters of the \220 lines.\label{tab2}}
\tablehead{
\colhead{Source} & \colhead{$\alpha$ [J2000]}\tablenotemark{1} & \colhead{$\delta$ [J2000]}\tablenotemark{1}
& \colhead{$v_{LSR}$}  & \colhead{$\Delta v$}  &   \colhead{$\int T_{\mathrm{MB}}(v)\,dv$} & \colhead{rms Noise} & \colhead{$N_{^{13}CO}$}\\ 
 \phn &  \colhead{[\ \fh \ \ \fm \ \ \fs]}  &  \colhead{[\degr \ \arcmin \
   \arcsec]} & \colhead{[\kms]} &  \colhead{[\kms]} &
   \colhead{[K \kms]} &  \colhead{[K]}& \colhead{[10$^{15}$~cm$^{-2}$]} } 
\startdata
DC259.5-16.4 & 07 14 03.2 & -48 27 40 &  4.6 & 1.5 $\pm$ 0.1 & 2.7 $\pm$ 0.2 & 0.1 & 1.1  \\  
DC253.8-10.9 & 07 29 32.7 & -41 07 30 &  -0.6 & 0.8 $\pm$ 0.1\tablenotemark{2} & 1.5 $\pm$ 0.2 & 0.2 & 0.6 \\  
DC255.4-3.9 & 08 05 26.3 & -39 07 26 &  9.8 & 2.4 $\pm$ 0.1 & 2.2 $\pm$ 0.2 & 0.2 & 0.9  \\  
CG31A-C & 08 08 41.8 & -35 54 51 & 6.7  & 1.9 $\pm$ 0.1 &  6.0$\pm$ 0.2 & 0.2 & 2.4 \\
$\star$ CG30 & 08 09 26.3 &  -36 02 21 & 6.5 & 1.2 $\pm$ 0.1\tablenotemark{2} & 4.0 $\pm$ 0.2 & 0.1 & 1.6 \\
DC257.3-2.5 & 08 17 17.1 & -39 47 55 &  8.8  & 2.0 $\pm$ 0.2 & 6.8 $\pm$ 0.5 & 0.6 & 2.7 \\  
DC266.0-7.5 & 08 21 22.9 & -49 47 47 & 4.4 & 1.7 $\pm$ 0.1  & 4.8 $\pm$ 0.2 &  0.1 & 1.9  \\
$\star$ DC267.4-7.5 & 08 25 48.3 & -50 59 48 & 5.8  & 1.7 $\pm$ 0.1  & 5.3 $\pm$ 0.2 & 0.2 & 2.1 \\  
DC267.5-7.4 & 08 26 46.5 & -51 04 01 & 6.5 & 1.5 $\pm$ 0.1  & 2.8 $\pm$ 0.2 & 0.2 & 2.9  \\  
DC253.6+2.9 & 08 28 44.0 &  -33 43 42  & 7.3  & 1.7 $\pm$ 0.1 & 4.7 $\pm$ 0.2 & 0.2 & 1.9 \\  
DC267.6-6.4 & 08 32 10.2 & -50 31 21  &  6.3 &1.6 $\pm$ 0.1 & 2.8 $\pm$ 0.2 &  0.1 & 1.1\\  
DC269.4+3.0 & 09 22 22.0 & -45 54 55 & -2.9  & 1.3 $\pm$ 0.2\tablenotemark{2} & 1.9 $\pm$ 0.1 & 0.1 & 0.9 \\  
DC274.2-0.4 & 09 28 47.0 & -51 35 10 &  6.4  & 2.2 $\pm$ 0.2 &   2.5 $\pm$ 0.2 & 0.2 & 1.0 \\  
DC272.5+2.0 & 09 31 02.8 & -48 36 52 &  -3.9 & 1.6 $\pm$ 0.2  & 1.6 $\pm$0.2 & 0.1 & 0.6 \\  
$\star$ DC275.9+1.9 & 09 46 36.5 & -51 00 42 & -4.9 & 2.9 $\pm$ 0.2 & 3.4 $\pm$ 0.2 & 0.1 & 1.3  \\  
DC291.0-3.5 & 11 00 10.1 & -63 40 43 &  -3.8  & 1.7 $\pm$ 0.1 & 2.9 $\pm$ 0.2 &  0.1  & 1.1\\  
DC291.1-1.7 & 11 07 04.8 & -62 03 43 & -4.2  & 1.8 $\pm$ 0.1 & 3.2 $\pm$ 0.2 & 0.1 & 1.3 \\  
DC297.7-2.8 & 12 01 57.0 & -65 05 19 & -3.9 & 2.1 $\pm$ 0.1 & 7.4 $\pm$ 0.2 & 0.2 & 2.9 \\
DC300.2-3.5 & 12 25 03.5 & -66 08 52  & -5.2 & 1.9 $\pm$ 0.2 & 2.3 $\pm$ 0.2 & 0.2 & 0.9 \\  
Mu8 & 12 30 30.5 & -71 03 35  & 3.8  & 1.7 $\pm$ 0.1  & 4.4 $\pm$ 0.2 & 0.1 & 1.7 \\  
DC300.7-1.0 & 12 31 20.5 &  -63 41 32  & -5.5  & 1.1 $\pm$ 0.2\tablenotemark{2} & 2.0 $\pm$ 0.1 & 0.2 & 0.8 \\  
DC301.2-0.4 & 12 36 01.7 &  -63 08 39 & -4.4  & 1.6 $\pm$ 0.1  & 1.8 $\pm$ 0.2  & 0.1 & 0.7 \\  
DC302.6-15.9 & 12 44 26.8 &  -78 45 21  &  4.1  & 1.8 $\pm$ 0.1 & 1.8 $\pm$ 0.2 & 0.1 & 0.7 \\  
$\star$ DC302.1+7.4 & 12 45 49.1 & -55 22 18 &  -14.7  & 2.1 $\pm$ 0.2  & 2.6 $\pm$ 0.2 & 0.1 & 1.0 \\  
$\star$ DC303.8-14.2 & 13 07 35.7 &  -76 57 05 &  4.3  & 2.1 $\pm$ 0.1 & 3.6 $\pm$ 0.2  & 0.1 & 1.4 \\  
DC327.2+1.8 & 15 42 49.0 & -52 49 39  & -0.4  & 2.0 $\pm$ 0.2  & 2.3 $\pm$ 0.2 & 0.1 & 0.9 \\  
DC339.1+11.7 & 15 59 11.2 & -37 36 57 & 5.7  & 1.7 $\pm$ 0.1 & 4.3 $\pm$ 0.2 & 0.1 & 1.7 \\  
DC338.6+9.5 & 16 04 43.6 & -39 36 26 & 5.2  & 1.0 $\pm$ 0.1\tablenotemark{2}  & 3.1 $\pm$ 0.2 & 0.1 & 1.2 \\  
DC337.6+7.6 & 16 08 21.0 &  -41 40 49 & 5.4  & 1.7 $\pm$ 0.1 & 7.1 $\pm$ 0.1 & 0.2 & 2.8 \\  
$\star$ L43 & 16 34 23.7 & -15 49 57 &  1.0  & 1.8 $\pm$ 0.1 & 6.6 $\pm$ 0.2  & 0.1 & 2.6 \\  
DC346.0+7.8 & 16 36 54.7 & -35 37 38 & 4.1  & 2.2 $\pm$ 0.1 & 5.4 $\pm$ 0.2 & 0.2 & 2.1 \\  
DC346.3+7.8 & 16 37 37.5 & -35 30 53 & 3.8  & 1.8 $\pm$ 0.1 & 4.7 $\pm$ 0.2 & 0.2 & 1.9 \\  
DC346.4+7.9 & 16 37 47.7 &  -35 14 11 &  4.1  & 1.7 $\pm$ 0.1 & 4.0 $\pm$ 0.2 & 0.2 & 1.6 \\  
L63 & 16 49 52.1 & -18 03 04 &  6.3  & 2.0 $\pm$ 0.1  & 7.3 $\pm$ 0.1  & 0.2 & 2.9 \\  
$\star$ CB68 & 16 56 49.3 &  -16 04 32 &  5.1  & 1.7 $\pm$ 0.1  & 3.2 $\pm$ 0.1  & 0.2 & 1.3 \\  
$\star$ B59 & 17 11 26.6  & -27 25 58  &  3.7  & 2.1 $\pm$ 0.1 & 10.9 $\pm$ 0.2  & 0.2 & 4.3 \\  
$\star$ L100 & 17 16 01.0  & -20 57 21 &  2.1 & 2.2 $\pm$ 0.1  & 4.1 $\pm$ 0.3 & 0.1 & 1.6 \\  
L1772 & 17 19 32.7 & -26 41 40 & 3.9  & 1.8 $\pm$ 0.1  & 3.5 $\pm$ 0.2 & 0.2 & 1.4 \\  
B72 & 17 23 38.5 &  -23 38 28 & 5.4  & 1.7 $\pm$ 0.1  & 3.4 $\pm$ 0.2 & 0.2 & 1.3 \\  
L328 & 18 16 53.7 & -18 03 24 & 7.4  & 2.7 $\pm$ 0.2  & 2.8  $\pm$ 0.2 &  0.2 & 1.1
\enddata
\tablenotetext{1}{\ Coordinates of spectrum with highest integrated intensity.}
\tablenotetext{2}{\ Linewidth obtained using high resolution spectrometer data.}
\end{deluxetable}

%%% Local Variables: 
%%% mode: latex
%%% TeX-master: t
%%% End: 

%% file: tab3.tex
\begin{deluxetable}{lrrrrrr}
\tablewidth{0pc}
\tablenum{3}
\tablecaption{Observed parameters of the \460 lines.\label{tab3}}
\tablehead{
\colhead{Source} & \colhead{$\alpha$ [J2000]}\tablenotemark{1} & \colhead{$\delta$ [J2000]}\tablenotemark{1}& 
 \colhead{$v_{LSR}$}  & \colhead{$\Delta v$} &  \colhead{$\int T_{\mathrm{MB}}(v)\,dv$} & rms Noise \\ 
 \phn &  \colhead{[\ \fh \ \ \fm \ \ \fs]}  &  \colhead{[\degr \ \arcmin \
   \arcsec]}&
 \colhead{[\kms]} &  \colhead{[\kms]} & \colhead{[K \kms]}& 
\colhead{[K]} } 
        \startdata
DC259.5-16.4 & 07 14 06.2 & -48 28 10 & 4.2 & 3.0 $\pm$ 0.4  & 3.9 $\pm$ 0.4 & 0.4  \\
DC253.8-10.9 &  07 29 32.7 & -41 10 00 & -1.1 & 3.1 $\pm$ 0.3  & 4.1 $\pm$ 0.3 &  0.3  \\
DC255.4-3.9  & 08 05 18.6 & -39 11 26 & 9.7 & 2.1 $\pm$ 0.3  & 3.2 $\pm$ 0.3 & 0.4 \\
CG31A-C & 08 09 01.5 & -35 59 21 & 6.7 & 2.2 $\pm$ 0.1 &  14.4 $\pm$ 0.5 & 0.3 \\
$\star$\tablenotemark{2} CG30 & 08 09 33.7 & -36 04 51 & 6.5 & 4.2 $\pm$ 0.3 &  11.0 $\pm$ 0.3 & 0.6 \\
DC257.3-2.5 & 08 17 11.9 & -39 51 25  &  8.6 & 1.4 $\pm$ 0.1 &  5.9 $\pm$ 0.4  & 0.4 \\
DC266.0-7.5 & 08 21 13.6  & -49 54 47 &  3.8  & 2.0 $\pm$ 0.2   & 4.9$\pm$ 0.3 & 0.3   \\
$\star$ DC267.4-7.5 & 08 25 45.1 & -50 59 18 &  5.5 & 1.8 $\pm$ 0.1 & 10.9 $\pm$ 0.2 & 0.2 \\
DC267.5-7.4 & 08 26 56.1 & -51 04 31 &  6.0  & 2.7 $\pm$ 0.2  &  5.9 $\pm$ 0.3  &  0.3  \\
DC253.6+2.9 & 08 28 58.4 & -33 47 12 & 6.3  & 1.3 $\pm$ 0.1  & 6.4 $\pm$ 0.3   & 0.3  \\
DC267.6-6.4 & 08 32 03.9  & -50 32 21  &  5.8  & 1.6 $\pm$ 0.1  &   5.2 $\pm$ 0.3  & 0.3 \\
DC269.4+3.0 & 09 22 16.3 & -45 47 25 &   -1.4 & 2.0 $\pm$ 0.2 & 3.3 $\pm$ 0.2  & 0.3 \\
DC274.2-0.4 & 09 28 50.2 & -51 36 40 & 5.9  & 2.2 $\pm$ 0.2  &  2.6 $\pm$ 0.2   & 0.2  \\
DC272.5+2.0 & 09 31 11.9 & -48 37 52 & -3.8 & 2.7 $\pm$ 0.9  & 1.5 $\pm$ 0.2   & 0.3 \\
$\star$ DC275.9+1.9 & 09 46 30.1 & -51 01 12 & -5.7  & 2.6 $\pm$ 0.3  &  4.2 $\pm$ 0.4  &  0.4 \\
DC291.0-3.5 & 10 59 02.4 & -63 40 43 &  -4.8 & 1.0 $\pm$ 0.1 & 5.0 $\pm$ 0.4 & 0.6 \\
DC291.1-1.7 & 11 06 13.6 & -62 03 13 &  -4.3  & 1.2 $\pm$ 0.1  & 3.5 $\pm$ 0.2 & 0.3 \\
DC300.2-3.5 & 12 24 19.0  & -66 12 52  &   -5.7  & 1.1 $\pm$ 0.1  & 2.7 $\pm$ 0.2 & 0.2  \\
Mu8 & 12 28 07.6 & -71 19 05 &  3.25 & 2.5 $\pm$ 0.3 & 6.4 $\pm$ 0.5 & 0.7 \\
DC300.7-1.0 & 12 31 29.5 & -63 44 32   & -4.9  & 3.5 $\pm$ 0.1  & 5.9 $\pm$ 0.1 & 1.3 \\
DC301.2-0.4 & 12 36 19.4  & -63 11 09  & -4.7  & 2.6 $\pm$ 0.5  &  2.4 $\pm$ 0.3  &  0.3 \\
DC302.6-15.9 & 12 45 38.7 & -78 47 21 &  3.8 & 1.7 $\pm$ 0.1 &  2.2 $\pm$ 0.1 & 0.5 \\
$\star$ DC302.1+7.4 & 12 45 38.5  & -55 24 48 & -14.7  &  2.2 $\pm$ 0.3 &  3.7 $\pm$ 0.4 & 0.4   \\
$\star$ DC303.8-14.2 & 13 07 26.8  & -77 01 35  & 3.4  & 5.8 $\pm$ 0.5  & 6.7 $\pm$ 0.3 &  0.2  \\
DC327.2+1.8 & 15 42 32.4 & -52 48 09 & -0.8  & 3.2 $\pm$ 0.5  &  3.1 $\pm$ 0.3  & 0.3 \\
DC339.1+11.7 & 15 59 31.4 & -37 36 57  &  5.1  & 1.3 $\pm$ 0.1  &  6.6 $\pm$ 0.5  & 0.5 \\
DC338.6+9.5 & 16 04 33.2 & -39 38 56 &   4.4 & 1.0 $\pm$ 0.1  &  3.6 $\pm$ 0.3  & 0.3  \\
DC337.6+7.6 & 16 07 46.2 & -41 43 49 &  4.7 & 1.7 $\pm$ 0.1 &  9.5 $\pm$ 0.1  & 0.5 \\
$\star$ L43 & 16 34 25.8  & -15 48 27   & 0.4  & 3.0 $\pm$ 0.3  & 14.8 $\pm$ 1.0 & 1.0 \\
DC346.0+7.8 & 16 36 57.2 & -35 36 38 & 3.3 & 1.8 $\pm$ 0.2 & 6.8 $\pm$ 0.4  & 0.4 \\
DC346.3+7.8 & 16 37 47.4 & -35 28 53 & 3.4  & 1.8 $\pm$ 0.1  &  7.1 $\pm$ 0.3  & 0.3 \\
DC346.4+7.9 & 16 37 57.5 & -35 15 11  & 3.3  & 1.6 $\pm$ 0.1  &  4.8 $\pm$ 0.2 &  0.3  \\
L63 & 16 50 04.7 & -18 01 04  & 5.8 & 1.2 $\pm$ 0.1  & 6.7 $\pm$ 0.2  & 1.4 \\
$\star$ CB68 & 16 57 03.8 & -16 03 32 &  5.6 & 3.4 $\pm$ 0.4 & 7.5 $\pm$ 0.4  & 1.2 \\
$\star$ B59 & 17 11 17.6 & -27 26 28  & 4.0 & 3.6 $\pm$ 0.3 &  18.1 $\pm$ 0.7  & 0.9 \\
$\star$ L100 & 17 16 16.0 & -21 01 51  & 1.4 & 1.8 $\pm$ 0.3 & 14.2 $\pm$ 1.6  & 0.4 \\
L1772 & 17 19 50.6 & -26 44 40 &  3.5  & 2.1 $\pm$ 0.2  &  5.1 $\pm$ 0.3   & 0.4 \\
B72 & 17 23 04.7 & -23 42 28 & 5.1 & 1.9 $\pm$ 0.3  &  6.0 $\pm$ 0.5   & 0.8 \\
L328 & 18 17 02.1  & -18 04 24 & 6.9 &  1.2 $\pm$ 0.3 &  3.3 $\pm$ 0.7  &  0.9 
\enddata
\tablenotetext{1}{\ Coordinates of spectrum with highest integrated intensity.}
\tablenotetext{2}{\ Cores with known embedded sources as detected using IRAS are
  marked ($\star$).}
\end{deluxetable}

%%% Local Variables: 
%%% mode: latex
%%% TeX-master: t
%%% End: 

%% file: tab4.tex
\begin{deluxetable}{lrrrrrrrrrrr}
\tablewidth{0pc}
\tablenum{4}
\rotate
\tablecaption{Results from the LVG model.}
\tablehead{
\colhead{Source} & \colhead{$\alpha$ [J2000]}\tablenotemark{1} & \colhead{$\delta$ [J2000]}\tablenotemark{1}
&  \colhead{T$^{13}_{2\rightarrow 1}$} & \colhead{T$^{12}_{4\rightarrow 3}$} &
\colhead{T$^{12}_{7\rightarrow 6}$} & \colhead{$\frac{T^{12}_{7\rightarrow
    6}}{T^{12}_{4\rightarrow 3}}$} & 
\colhead{$\frac{T^{12}_{4\rightarrow 3}}{T^{13}_{2\rightarrow 1}}$} &
\colhead{T$_{kin}$} & \colhead{T$_{kin}$} & \colhead{log n(H$_{2}$)} &  \colhead{log n(H$_{2}$)} \\ 
 \phn &  \colhead{[\ \fh \ \ \fm \ \ \fs]}  &  \colhead{[\degr \ \arcmin \
   \arcsec]}  & \colhead{[K]} & \colhead{[K]} & \colhead{[K]} & \phn & \phn &
 \colhead{[K]} & \colhead{[K]}  & \colhead{[cm$^{-3}$]} & \colhead{[cm$^{-3}$]} \\ 
\phn & \phn & \phn & \phn & \phn & \colhead{u.l.\tablenotemark{3}} &
\colhead{u.l.\tablenotemark{3}} & \phn & \colhead{u.l.\tablenotemark{3}}
&\colhead{l.l.\tablenotemark{3}} & \colhead{u.l.\tablenotemark{3}} &
\colhead{l.l.\tablenotemark{3}} \label{tab4} }
\startdata
DC259.5-16.4 & 07 14 03.2 & -48 27 40 & 1.6 & 0.8 & 0.5 & 0.6 & 0.5 & 22 &  14
& 3.2 & 2.9 \\
CG31A-C & 08 08 49.2 & -35 57 51 & 3.1 & 4.3 & 1.4 & 0.3 & 1.4 & 28 & 17 &
3.2 & 2.8  \\
$\star$\tablenotemark{2} CG30 & 08 09 26.3 & -36 03 51 & 1.6 & 1.8 & 0.8 & 0.4
& 1.2 & 30 & 18 & 3.3 & 2.9 \\
$\star$ DC267.4-7.5 & 08 25 48.3 & -50 59 48 & 2.5 & 2.8 & 0.3 & 0.1 & 1.1 &
22 & 14 & 3.2 & 2.9 \\
DC267.5-7.4 & 08 26 46.5 & -51 04 01 & 1.6 & 1.5 & 0.4 & 0.3 & 0.9 & 26 & 15 &
3.2 & 2.9 \\
DC274.2-0.4 & 09 28 47.0 & -51 35 10 & 1.0 & 0.6 & 0.2 & 0.4 & 0.6 & 18 & 12 &
3.7 & 3.0 \\
DC339.1+11.7 & 15 59 11.2 & -37 36 57 & 2.3 & 3.4 & 1.3 & 0.4 & 1.5 & 34 & 19
& 3.1 & 2.8 \\
$\star$ B59 & 17 11 19.8 & -27 25 58 & 4.0 & 4.1 & 1.9 & 0.5 & 1.0 & 28 & 16 &
3.3 & 2.9\\
$\star$ L100 & 17 16 07.4 & -20 57 21 & 1.7 & 2.3 & 0.6 & 0.3 & 1.4 & 30 & 20
& 3.1 & 2.7
\enddata
\tablenotetext{1}{\ Coordinates of position where LVG analysis was performed.}
\tablenotetext{2}{\ Cores with known embedded sources as detected using IRAS are
  marked ($\star$).}
\tablenotetext{3}{\ Upper and lower limits are denoted u.l. and l.l.}
\end{deluxetable}

%%% Local Variables: 
%%% mode: latex
%%% TeX-master: t
%%% End: 

%% file: tab5.tex
\begin{deluxetable}{lrrrrrr}
\tablewidth{0pc}
\tablenum{5}
\tablecaption{Masses derived from the H$_{2}$ column densities.}
\tablehead{
\colhead{Source} & \colhead{Distance}\tablenotemark{1} & \colhead{Radius$_{1}$}\tablenotemark{2} & \colhead{Radius$_{2}$}\tablenotemark{2} & 
\colhead{N(H$_{2}$)} & \colhead{Mass} & \colhead{T$_{kin}$} \\
 \phn & \colhead{[pc]} & \colhead{[pc]} &  \colhead{[pc]} & 
\colhead{[10$^{21}$~cm$^{-2}$]} & \colhead{[M$_{\odot}$]} & \colhead{u.l.\tablenotemark{3} [K]} \label{tab5}} 
 \startdata
DC259.5-16.4 & 450 $\pm$ 50 & 0.9 & 0.7 & 1.3 & 48 $\pm$ 12 & 22 \\
$\star$\tablenotemark{4} CG30 & 450 $\pm$ 50 & 0.8 & 0.7 & 1.8 & 58 $\pm$ 14 & 30 \\
$\star$ DC267.4-7.5 & 450 $\pm$ 50 & 1.4  & 1.2  & 2.9 & 255 $\pm$ 60 & 22 \\
DC253.6+2.0 & 450 $\pm$ 50 & 0.9 & 0.7 & 2.4 & 103 $\pm$ 26 & \nodata \\
DC274.2-0.4 & 450 $\pm$ 50 & 0.8 & 0.7 & 1.5 & 56 $\pm$ 14 & 18 \\
DC297.7-2.8 & 150 $\pm$ 30 & 0.2 & 0.3 & 3.3 & 12 $\pm$ 4 & \nodata \\
DC302.6-15.9 & 150 $\pm$ 30 & 0.2 & 0.3 & 0.9 &  4 $\pm$ 2 & \nodata \\
$\star$ DC303.8-14.2 & 178 $\pm$ 18 & 0.5 & 0.4 & 1.5 & 19 $\pm$ 4 & \nodata \\
$\star$ L100 & 125 $\pm$ 25 & 0.4 & 0.4 & 1.9 & 14 $\pm$ 6  & 30  
\enddata
\tablenotetext{1}{\ Distances as derived in
  \citet{woermann,franco,corradi,whittet,knude, geus}}
\tablenotetext{2}{\ Radii of the elliptical half power width of the core.} 
\tablenotetext{3}{\ Upper limit as derived from the LVG analysis.}
\tablenotetext{4}{\ Cores with known embedded sources as detected using IRAS are
  marked ($\star$).}
\end{deluxetable}
%%% Local Variables: 
%%% mode: latex
%%% TeX-master: t
%%% End: 

%% file: ms.bbl
\begin{thebibliography}{}

\bibitem[Balakrishnan et al. (2002)]{balakrishna02}
        Balakrishnan, N., Yan, M., \& Dalgarno, A. 2002, \apj, 568, 443 

\bibitem[Bourke et al. (2005)]{tyla2005}
Bourke, T. L., Crapsi, A., Myers, P. C., Evans, N. J., Wilner, D. J., Huard,
T. L., J{\o}rgensen, J. K., \& Young, C. H. 2005, \apjl, 633, L129

\bibitem[Bourke et al. (1995)]{tyla95}
        Bourke, T. L., Hyland, A. R., \& Robinson, G. 1995, \mnras, 276, 1052

\bibitem[Bourke et al. (1997)]{tyla97}
        Bourke, T. L., et al. 1997, \apj, 467, 781

\bibitem[Bourke et al. (2006)]{tyla2006} 
Bourke, T. L., et al. 2006, \apjl, 649, L37

\bibitem[Brooke et al. (2007)]{brooke}
        Brooke, T. Y. et al., 2007, \apj, 655, 364

\bibitem[Chamberlin et al. (1997)]{cls1997}
  Chamberlin, R. A., Lane, A. P., \& Stark, A. A. 1997, \apj, 476, 428

\bibitem[Corradi et al. (1997)]{corradi}
        Corradi, W. J. B., Franco, G. A. P., \& Knude, J. 1997, \aap, 216, 44

\bibitem[Dunham et al. (2006)]{dunham2006} 
Dunham, M. M., et al. 2006, \apj, 651, 945

\bibitem[Evans et al. (2003)]{evans03}
        Evans, N. J., et al. 2003, \pasp, 115, 965

\bibitem[di Francesco et al. (2006)]{difrancesco}
        Di Francesco, J., Evans, N. J., Caselli, P., Muers, P., C., Shirley, Y., Aikawa, Y., \& Tafalla, M. 2006,
 Protostars and Planets V review, (astro-ph/0602379)

\bibitem[Franco (1995)]{franco}
        Franco, G. A. P. 1995, A\&AS, 114, 105
        
\bibitem[Frerking et al. (1982)]{frerk82}
        Frerking, M. A., Langer, D. W., \& Wilson, R. W. 1982, \apj, 262, 590

\bibitem[de Geus et al. (1989)]{geus}
        de Geus, E. J., de Zeeuw, P. T., \& Lub, J. 1989, \aap, 216, 44

\bibitem[Goldreich \& Kwan (1974)]{goldr74}
Goldreich, P., \& Kwan, J. 1974, \apj, 189, 441

\bibitem[Honingh et al. (1997)]{hon1997}
  Honingh, C. E., Haas, S., Hottgenroth, D., Jacobs, K., \& Stutzki, J. 1997, in
  8th Int. Symp. on Space Terahertz Technology, eds. R. Blundell \& E. Tong
  (Cambridge: Harvard Univ.), 92

\bibitem[Huard et al. (2006)]{huard2006}
        Huard, T. L., et al. 2006, \apj, 640, 391

\bibitem[Keto \& Myers (1986)]{keto86}
        Keto, E. R., \& Myers, P. C. 1986, \apj, 304, 466

\bibitem[Kim et al. (2002)]{skim}
        Kim, S., Martin, C. L., Stark, A. A., \& Lane, A. P. 2002, \apj, 580, 896

\bibitem[Knude \& H\o g (1998)]{knude}
        Knude, J., \& H\o g 1998, \aap, 338, 897

\bibitem[Kooi et al. (1992)]{kooi1992}
Kooi, J. W., Man, C., Phillips, T. G., Bumble, B., \& LeDuc, H. G. 1992, IEEE
Trans. Microwave Theory and Techniques, 40, 812

\bibitem[Lane (1998)]{apl1998}
  Lane, A. P. 1998, in ASP Conf. Ser. 141, Astrophysics From Antarctica,
  eds. G. Novack \& R. H. Landsberg (San Francisco: ASP), 289

\bibitem[Lee \& Myers (1999)]{lee99}
        Lee, C. W., \& Myers, P. C. 1999, \apjs, 123, 233

\bibitem[McGlynn et al. (1996)]{skyview}
McGlynn, T., Scollick, K., \& White, N. 1996, SkyView:
     The Multi-Wavelength Sky on the Internet, McLean, B.J.
     et al., New Horizons from Multi-Wavelength
     Sky Surveys, Kluwer Academic Publishers, IAU Symposium No. 179, p465.

\bibitem[Myers et al. (1987)]{myers1987} 
  Myers, P. C., Fuller, G. A., Mathieu, R. D., Beichman, C. A., Benson, P. J.,
 Schild, R. E., \& Emerson, J. P.\ 1987, \apj, 319, 340

\bibitem[Myers et al. (1983)]{myers83}
        Myers, P. C., Linke, R. A., \& Benson, P. J. 1983, \apj, 264, 517

\bibitem[Nielsen et al. (1998)]{nielsen98}
        Nielsen, A. S., Olberg, M., Knude, J., \& Booth, R. S. 1998, \aap, 336, 329

\bibitem[Oberst et al. (2006)]{oberst06}
        Oberst, T. E., et al. 2006, \apjl, 652, L1250

\bibitem[Otrupcek et al.(2000)]{otru2000} 
Otrupcek, R. E., Hartley, M., \& Wang, J.-S. 2000, Publ. Ast. Soc. of Australia, 17, 92

\bibitem[Schieder et al. (1989)]{stw1989}
  Schieder, R., Tolls, V., \& Winnewisser, G. 1989, Exp. Astron., 1, 101

\bibitem[Stark et al. (2001)]{aas2001}
  Stark, A. A., et al. 2001, \pasp, 113, 567

\bibitem[Tafalla et al. (2002)]{tafalla}
        Tafalla, M., Myers, P. C., Caselli, P., Walmsley, C. M., \& Comito,
        C. 2002, \apj, 569

\bibitem[Turner (1995)]{turner95}
        Turner, B. E. 1995, \apj, 455, 556

\bibitem[Vilas-Boas et al. (1994)]{vilas94}
         Vilas-Boas, J. W. S., Myers, P. C., \& Fuller, G. A. 1994, \apj, 433,
         96

\bibitem[Vilas-Boas et al. (2000)]{vilas2000}
         Vilas-Boas, J. W. S., Myers, P. C., \& Fuller, G. A. 2000, \apj, 532, 1038

\bibitem[Walker et al. (1992)]{w1992}
  Walker, C. K., Kooi, J. W., Chan, M., Leduc, H. G., Schaffer, P. L.,
  Carlstrom, J. E., \& Phillips, T. G. 1992, Int. J. Infrared MM Waves,
  13, 785 

\bibitem[Ward-Thompson et al. (2006)]{ward}
        Ward-Thompson, D., Andr$\acute{e}$, P., Crutcher, R., Johnstone, D., Onishi, T., \& Wilson, C. 2006, Protostars and Planets V review, (astro-ph/0603475)

\bibitem[Wilking (1992)]{wilki92}
  Wilking, B. A. 1992, in {\it{Low Mass Star Formation in Southern Molecular
  Clouds}}, ESO Scientific Report No. 11, ed. B. Reipurth, (ESO: Garching), p.159

\bibitem[Whittet et al. (1997)]{whittet}
        Whittet, D. C. B., Prusti, T., Franco, G. A. P. Gerakines, P. A., 
        Kilkenny, D., Larson, K. A. \& Wesselius, P. R. 1997, \aap, 327, 1194

\bibitem[Wilson \& Rood (1994)]{tomwilson94}
        Wilson, T. L., \& Rood, R. T. 1994, \araa, 191

\bibitem[Woermann et al. (2001)]{woermann}
        Woermann, B., Gaylard, M. J., \& Otrupcek, R. 2001, \mnras, 325, 1213

\bibitem[Young et al. (2004)]{young2004} 
  Young, C. H., et al. 2004, \apjs, 154, 396

\end{thebibliography}
